\documentclass[preprint]{aastex}
\doublespace
\usepackage{graphicx}
\shorttitle{IPN Supplement}
\shortauthors{Pal'shin et al.}

\begin{document}

\title{IPN localizations of Konus short gamma-ray bursts}

\author{V. D. Pal'shin\altaffilmark{1}, K. Hurley\altaffilmark{2}, D. S. Svinkin\altaffilmark{1}, %
R. L. Aptekar\altaffilmark{1}, S. V. Golenetskii\altaffilmark{1}, %
D. D. Frederiks\altaffilmark{1}, E. P. Mazets\altaffilmark{1,3}, %
P. P. Oleynik\altaffilmark{1}, M. V. Ulanov\altaffilmark{1}, T. Cline\altaffilmark{4,29}, %
I. G. Mitrofanov\altaffilmark{5}, D. V. Golovin\altaffilmark{5}, %
A. S. Kozyrev\altaffilmark{5}, M. L. Litvak\altaffilmark{5}, A. B. Sanin\altaffilmark{5}, %
W. Boynton\altaffilmark{6}, C. Fellows\altaffilmark{6}, K. Harshman\altaffilmark{6}, %
J. Trombka\altaffilmark{4}, T. McClanahan\altaffilmark{4}, R. Starr\altaffilmark{4}, %
J. Goldsten\altaffilmark{7}, R. Gold\altaffilmark{7}, %
A. Rau\altaffilmark{8}, A. von Kienlin\altaffilmark{8}, V. Savchenko\altaffilmark{9}, %
D. M. Smith\altaffilmark{10}, W. Hajdas\altaffilmark{11}, %
S. D. Barthelmy\altaffilmark{4}, J. Cummings\altaffilmark{12,30}, %
N. Gehrels\altaffilmark{4}, H. Krimm\altaffilmark{12,31}, D. Palmer\altaffilmark{13}, %
K. Yamaoka\altaffilmark{14}, M. Ohno\altaffilmark{15}, %
Y. Fukazawa\altaffilmark{15}, Y. Hanabata\altaffilmark{15}, %
T. Takahashi\altaffilmark{14}, M. Tashiro\altaffilmark{16}, %
Y. Terada\altaffilmark{16}, T. Murakami\altaffilmark{17}, K. Makishima\altaffilmark{18}, %
M. S. Briggs\altaffilmark{19}, R. M. Kippen\altaffilmark{13}, %
C. Kouveliotou\altaffilmark{20}, C. Meegan\altaffilmark{21}, G. Fishman\altaffilmark{20}, %
V. Connaughton\altaffilmark{19}, M. Bo\"{e}r\altaffilmark{22}, %
C. Guidorzi\altaffilmark{23}, F. Frontera\altaffilmark{23,32}, %
E. Montanari\altaffilmark{23,33}, F. Rossi\altaffilmark{23}, %
M. Feroci\altaffilmark{24}, L. Amati\altaffilmark{25}, %
L. Nicastro\altaffilmark{25}, M. Orlandini\altaffilmark{25}, %
%
E. Del Monte\altaffilmark{24}, E. Costa\altaffilmark{24}, %
I. Donnarumma\altaffilmark{24}, Y. Evangelista\altaffilmark{24}, %
I. Lapshov\altaffilmark{24}, F. Lazzarotto\altaffilmark{24}, %
L. Pacciani\altaffilmark{24}, M. Rapisarda\altaffilmark{24}, %
P. Soffitta\altaffilmark{24}, %
G. Di Cocco\altaffilmark{25}, F.~Fuschino\altaffilmark{25}, %
M. Galli\altaffilmark{26}, C. Labanti\altaffilmark{25}, M. Marisaldi\altaffilmark{25}, %
J.-L. Atteia\altaffilmark{27}, R. Vanderspek\altaffilmark{28}, %
G. Ricker\altaffilmark{28}}
\altaffiltext{1}{Ioffe Physical Technical Institute, St. Petersburg,
194021, Russian Federation \email{val@mail.ioffe.ru}}
\altaffiltext{2}{Space Sciences Laboratory, University of
California, 7 Gauss Way, Berkeley, CA 94720-7450, USA}
\altaffiltext{3}{Deceased}
\altaffiltext{4}{NASA Goddard Space Flight Center, Greenbelt, MD
20771, USA}
\altaffiltext{5}{Space Research Institute, 84/32, Profsoyuznaya,
Moscow 117997, Russian Federation}
\altaffiltext{6}{Department of Planetary Sciences, University of
Arizona, Tucson, Arizona 85721, USA}
\altaffiltext{7}{Applied Physics Laboratory, Johns Hopkins
University, Laurel, MD 20723, USA}
\altaffiltext{8}{Max-Planck-Institut f\"{u}r extraterrestrische
Physik, Giessenbachstrasse, Postfach 1312, Garching, 85748 Germany}
\altaffiltext{9}{Fran\c{c}ois Arago Centre, APC, Universit\'e Paris
Diderot, CNRS/IN2P3, CEA/Irfu, Observatoire de Paris, Sorbonne Paris
Cit\'e, 10 rue Alice Domon et L\'eonie Duquet, 75205 Paris Cedex 13,
France}
\altaffiltext{10}{Physics Department and Santa Cruz Institute for
Particle Physics, University of California, Santa Cruz, Santa Cruz,
CA 95064, USA}
\altaffiltext{11}{Paul Scherrer Institute, 5232 Villigen PSI,
Switzerland}
%
\altaffiltext{12}{UMBC/CRESST/NASA Goddard Space Flight Center,
Greenbelt, MD 20771, USA}
\altaffiltext{13}{Los Alamos National Laboratory, Los Alamos, New
Mexico 87545, USA}
%
\altaffiltext{14}{Institute of Space and Astronautical Science
(ISAS/JAXA), 3-1-1 Yoshinodai, Chuo-ku, Sagamihara, Kanagawa
252-5210, Japan}
\altaffiltext{15}{Department of Physics, Hiroshima University, 1-3-1
Kagamiyama, Higashi-Hiroshima, Hiroshima 739-8526, Japan}
\altaffiltext{16}{Department of Physics, Saitama University, 255
Shimo-Okubo, Sakura-ku, Saitama-shi, Saitama 338-8570, Japan}
\altaffiltext{17}{Department of Physics, Kanazawa University,
Kadoma-cho, Kanazawa, Ishikawa 920-1192, Japan}
\altaffiltext{18}{Department of Physics, University of Tokyo, 7-3-1
Hongo, Bunkyo-ku, Tokyo 113-0033, Japan}
%
\altaffiltext{19}{CSPAR and Physics Department, University of
Alabama in Huntsville, Huntsville, AL 35899, USA}
\altaffiltext{20}{Space Science Office, VP62, NASA Marshall Space
Flight Center, Huntsville, AL 35812, USA}
%
\altaffiltext{21}{Universities Space Research Association, 320
Sparkman drive, Huntsville, AL 35805}
\altaffiltext{22}{Observatoire de Haute Provence (CNRS), 04870 Saint
Michel l'Observatoire, France}
%
\altaffiltext{23}{Physics Department, University of Ferrara, Via
Saragat, 1, 44100 Ferrara, Italy}
\altaffiltext{24}{INAF - Istituto di Astrofisica Spaziale e Fisica
Cosmica, via Fosso del Cavaliere, Rome, I-00133, Italy}
\altaffiltext{25}{INAF - Istituto di Astrofisica Spaziale e Fisica
Cosmica di Bologna, via Gobetti 101, I-40129 Bologna, Italy}
\altaffiltext{26}{ENEA-Bologna, Via Martiri Montesole, 4 I-40129
Bologna, Italy}
%
\altaffiltext{27}{Universit\'{e} de Toulouse; UPS-OMP; CNRS; IRAP;
14, avenue Edouard Belin, F-31400 Toulouse, France}
\altaffiltext{28}{Kavli Institute for Astrophysics and Space
Research, Massachusetts Institute of Technology, 70 Vassar Street,
Cambridge, MA 02139, USA.}
%
%
\altaffiltext{29}{Emeritus}
\altaffiltext{30}{UMBC Physics Department 1000 Hilltop Circle
Baltimore, MD 21250, USA}
\altaffiltext{31}{Universities Space Research Association, 10211
Wincopin Circle, Suite 500, Columbia, MD 20144, USA}
\altaffiltext{32}{INAF/Istituto di Astrofisica Spaziale e Fisica
Cosmica di Bologna, via Gobetti 101, 40129 Bologna, Italy}
\altaffiltext{33}{Istituto IS Calvi, Finale Emilia (MO), Italy}

\begin{abstract}
Between the launch of the \textit{GGS Wind} spacecraft in 1994
November and the end of 2010, the Konus-\textit{Wind} experiment
detected 296 short-duration gamma-ray bursts (including 23 bursts
which can be classified as short bursts with extended emission).
During this period, the IPN consisted of up to eleven spacecraft,
and using triangulation, the localizations of 271 bursts were
obtained. We present the most comprehensive IPN localization data on
these events. The short burst detection rate, $\sim$18 per year,
exceeds that of many individual experiments.
\end{abstract}



\section{INTRODUCTION}

Between 1994 November and 2010 December, the Konus gamma-ray
spectrometer aboard the \textit{Global Geospace Science Wind}
spacecraft detected 1989 cosmic gamma-ray bursts (GRBs) in the
triggered mode, 296 of which were classified as short-duration
gamma-ray bursts or short bursts with extended emission (EE). The
classification was made based on the duration distribution of an
unbiased sample of 1168 Konus-\textit{Wind} GRBs. The instrument
trigger criteria cause undersampling of faint short bursts relative
to faint long bursts, so this subsample of fairly bright (in terms
of peak count rate in Konus-\textit{Wind}'s trigger energy band)
bursts has been chosen for the purpose of classification. Taking in
account other characteristics of these short-duration bursts such as
hardness ratio and spectral lag shows that about 16\% of them can be
in fact Type II (collapsar-origin), or at least their classification
as Type I (merger-origin) is questionable (see \citealt{zhang09} for
more information on the Type I/II classification scheme).
Nevertheless we consider here all 296 Konus-\textit{Wind}
short-duration  and possible short-duration with EE bursts
(hereafter we refer to them simply as Konus short bursts). Full
details of the Konus-\textit{Wind} GRB classification are given in
Svinkin et al. (2013, in preparation).

Every short burst detected by Konus was searched for in the data of
the spacecraft comprising the interplanetary network (IPN). We found
that 271 ($\sim$92\%) of the Konus-\textit{Wind} short GRBs were
observed by at least one other IPN spacecraft, enabling their
localizations to be constrained by triangulation.

The IPN contained between 3 and 11 spacecraft during this period.
They were, in addition to Konus-\textit{Wind}: \textit{Ulysses} (the
solar X-ray/cosmic gamma-ray burst instrument, GRB), in heliocentric
orbit at distances between 670 and 3180 lt-s from Earth
\citep{hurley92}; the \textit{Near-Earth Asteroid Rendezvous}
mission (NEAR) \citep[the remote sensing X-ray/Gamma-Ray
Spectrometer, XGRS;][]{trombka99}, at distances up to 1300 lt-s from
Earth; \textit{Mars Odyssey} \citep[the Gamma-Ray Spectrometer (GRS)
that includes two detectors with GRB detection capabilities, the
gamma sensor head (GSH), and the High Energy Neutron Detector
(HEND);][]{boynton04,hurley06}, launched in 2001 April and in orbit
around Mars starting in 2001 October, up to 1250 lt-s from Earth
\citep{saunders04}; \textit{Mercury Surface, Space Environment,
Geochemistry, and Ranging} mission (\textit{MESSENGER}) \citep[the
Gamma-Ray and Neutron Spectrometer, GRNS;][]{goldsten07}, en route
to Mercury (in Mercury orbit since March 2011), launched in 2004
August, but commencing full operation only in 2007, up to $\sim$700
lt-s from Earth \citep{gold01,solomon07}; the \textit{International
Gamma-Ray Astrophysics Laboratory} (\textit{INTEGRAL}) \citep[the
anti-coincidence shield ACS of the spectrometer SPI,
SPI-ACS;][]{rau05}, in an eccentric Earth orbit at up to 0.5 lt-s
from Earth; and in low Earth orbits: the \textit{Compton Gamma-Ray
Observatory} \citep[the Burst and Transient Source Experiment,
BATSE;][]{fishman92}; \textit{BeppoSAX} \citep[the Gamma-Ray Burst
Monitor, GRBM;][]{frontera97, feroci97}; the \textit{Ramaty High
Energy Solar Spectroscopic Imager}
\citep[\textit{RHESSI};][]{lin02,smith02}; the \textit{High Energy
Transient Explorer} (\textit{HETE-2})
 \citep[the French Gamma-Ray Telescope, FREGATE;][]{ricker03,atteia03}; the \textit{Swift} mission \citep[the Burst Alert Telescope,
 BAT;][]{barthelmy05,gehrels04}; the \textit{Suzaku} mission \citep[the Wide-band All-sky Monitor, WAM;][]{yamaoka09,takahashi07};
 \textit{AGILE} (the Mini-Calorimeter, MCAL, and Super-AGILE) \citep{tavani09}; the \textit{Fermi} mission \citep[the Gamma-Ray Burst
Monitor, GBM;][]{meegan09}, the \textit{Coronas-F} solar observatory
(Helicon) \citep{oraevskii02}, the \textit{Cosmos 2326}
\citep[Konus-A;][]{aptekar98}, \textit{Cosmos 2367} (Konus-A2), and
\textit{Cosmos 2421}(Konus-A3) spacecraft, and the
\textit{Coronas-Photon} solar observatory (Konus-RF).

At least two other spacecraft detected GRBs during this period,
although they were not used for triangulation and therefore were
not, strictly speaking, part of the IPN.  They are the
\textit{Defense Meteorological Satellite Program} (DMSP)
\citep{terrell96, terrell98, terrell04} and the \textit{Stretched
Rohini Satellite Series} (SROSS) \citep{marar94}.

Here we present the localization data obtained by the IPN for 271
Konus-\textit{Wind} short bursts observed by at least one other IPN
s/c. In a companion paper, we present the durations, energy spectra,
peak fluxes, and fluences of these bursts.

\section{OBSERVATIONS}
For each Konus short gamma-ray burst, a search was initiated in the
data of the IPN spacecraft. For the near-Earth spacecraft and
\textit{INTEGRAL}, the search window was centered on the
Konus-\textit{Wind} trigger time, and its duration was somewhat
greater than the \textit{Wind} distance from Earth. For the
spacecraft at interplanetary distances, the search window was twice
the light-travel time to the spacecraft if the event arrival
direction was unknown, which was the case for most events.  If the
arrival direction was known, even coarsely, the search window was
defined by calculating the expected arrival time at the spacecraft,
and searching in a window around it.

\begin{figure}
\centering
\includegraphics[width=\textwidth]{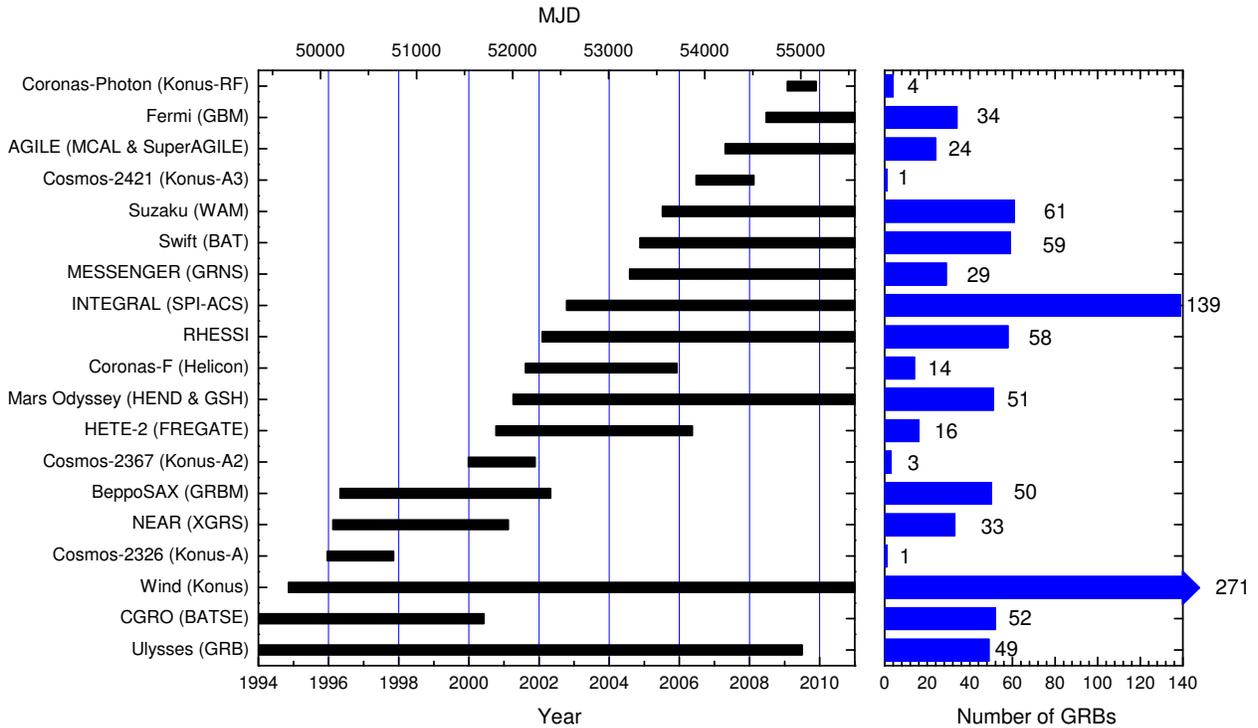}
\caption{\textit{Left}: Timelines of the IPN missions since the
launch of \textit{Wind} in 1994, November (instrument names are
given in the parentheses). \textit{Right:} Number of Konus short
bursts observed by each mission (for \textit{Wind} (Konus) -- the
number of bursts observed by at least one other IPN s/c is given).}
\label{Fig_TimeLines}
\end{figure}

The mission timelines and the number of Konus-\textit{Wind} short
GRBs observed by each mission/instrument are shown in
Figure~\ref{Fig_TimeLines}. In this study, the largest number of
bursts detected by an IPN instrument, after Konus, was 139, detected
by \textit{INTEGRAL} (SPI-ACS).

Table~\ref{Table_Basic} lists the 271 Konus-\textit{Wind} short GRBs
observed by the IPN. The first column gives the burst designation,
`\textsf{GRB}\texttt{YYYYMMDD}\textsf{\_T}\texttt{sssss}', where
\texttt{YYYYMMDD} is the burst date, and \texttt{sssss} is the
Konus-\textit{Wind} trigger time (s UT) truncated to integer seconds
(Note that, due to \textit{Wind}'s large distance from Earth, this
trigger time can differ by up to 6.1 seconds from the Earth-crossing
time; see below). The next two columns give the burst date and
Konus-\textit{Wind} trigger time in the standard date and time
formats. The `Type' column specifies the burst type following the
classification given in Svinkin et al. (2013, in preparation). The
types are: I (merger-origin), II (collapsar-origin), I/II (the type
is uncertain),  Iee (type I which shows extended emission), and
Iee/II (the type is uncertain: Iee or II). The `Time delay' column
gives the propagation time delay from \textit{Wind} to the Earth
center and its 3$\sigma$ uncertainty (calculated using the IPN
localizations, presented in this catalog: see section
\ref{Sec_ErrorRegions}). The `Observed by' column lists the
missions/instruments which observed the burst (detections by several
instruments which are not part of the IPN have also been listed,
namely COMPTEL on \textit{CGRO}, DMSP, \textit{Fermi} LAT, MAXI,
(\textit{Monitor of All-sky X-ray Image}), and SROSS
(\textit{Stretched Rohini Satellite Series}). The next two columns
give the total number of IPN s/c and the number of the distant IPN
s/c which observed the burst. The last column contains the comments.

During the period of consideration, four interplanetary s/c
participated in the IPN: \textit{Ulysses}, \textit{NEAR},
\textit{Mars Odyssey}, and \textit{MESSENGER}. Of the 271 bursts
listed in Table~\ref{Table_Basic}, 30 bursts were observed by two
distant s/c, 102 by one distant s/c, and 139 were not observed by
any distant s/c.

17 Konus short bursts were precisely localized by instruments with
imaging capabilities, namely, \textit{Swift}-BAT, \textit{HETE-2}
(WXM and SXC), and \textit{INTEGRAL} IBIS/ISGRI. For most of these
bursts an X-ray afterglow has been detected; for some of them a
redshift z has been determined based on the optical afterglow or
host galaxy spectroscopy. We have used these bursts to verify our
IPN triangulations (see section~\ref{Sec_AnnVerification}).

\section{METHODOLOGY}
When a GRB arrives at two spacecraft with a delay $\delta$T, it may
be localized to an annulus whose half-angle $\theta$ with respect to
the vector joining the two spacecraft is given by
\begin{equation}
\cos \theta=\frac{c \delta T}{D}
\end{equation}
where $c$ is the speed of light and $D$ is the distance between the
two spacecraft.  (This assumes that the burst is a plane wave, i.e.,
that its distance is much greater than $D$.)

The measured time delay has an uncertainty which is generally not
symmetrical $d_{\pm} (\delta T)$, i.e., the measured time delay can
take values from $\delta T + d_{-}(\delta T)$ to $\delta T + d_{+}
(\delta T)$ ($d_{-}(\delta T)$ is negative) at a given confidence
level.

The annulus half-widths $d \theta_\pm$, are
\begin{equation}
\label{EqHW}
d \theta_\pm \equiv \theta_\pm - \theta =  \cos^{-1} \left[\frac{c
(\delta T + d_\mp (\delta T))}{D} \right] - \cos^{-1} \left[\frac{c
\delta T }{D} \right]
\end{equation}

It should be noted that even in case of symmetrical errors
$|d_{-}(\delta T)|$ = $d_{+}(\delta T)$, the annulus can still be
significantly asymmetrical if $c(\delta T + d_\pm (\delta T))/D \sim
1$ (i.e. the source is close to the vector joining the two
spacecraft).

For the case $d (\delta T) \ll D/c$  Eq.~(\ref{EqHW}) reduces to
commonly used expression:
\begin{equation}
\label{EqHWsimple}
d \theta_\pm = -\frac{c d_\mp(\delta T)}{D \sin \theta}
\end{equation}

To derive the most probable time delay $\delta T$ and its
uncertainty $d_\pm (\delta T)$ we have used the $\chi^2$ method
described in~\citet{hurley99a} for triangulations with distant s/c
and this method with some modifications for
Konus-\textit{Wind}-near-Earth s/c (or \textit{INTEGRAL})
triangulations.

Given a burst time history recorded by two instruments, the most
probable time lag $\tau$  and its uncertainty $d_\pm(\tau)$ can be
estimated as follows. Let $n_{1,i} = n (t_{1,i})$, $n_{2,j} = n
(t_{2,j})$ and $\sigma_{1,i}$, $\sigma_{2,j}$ denote
background-subtracted counts and their uncertainties measured by two
instruments at evenly spaced intervals $t_{1,i} = t_{01} + i
\Delta_1$, $t_{2,j} = t_{02} + j \Delta_2$, where $i = 0,..,m_1$,
$j=0,..,m_2$; and $\Delta_1$, $\Delta_2$ are the bin sizes and
$t_{01}$, $t_{02}$ are the absolute reference times (UT). To make
things simpler, suppose $\Delta_1$ = $\Delta_2$ = $\Delta$. Usually
one assumes Poisson statistics, so $\sigma_{1(2),i} =
n_{tot1(2),i}^{1/2}$, where $n_{tot1(2),i}$ is the total number of
counts (source + background) in the $i$-th bin. Let us assume that
both time histories contain the burst of interest along with some
intervals before and after it (if they do not exist they can always
be added and padded with zeros with the background variance), and
$N+1$ bins from $i_{start}$ contain the burst (or the part of the
burst we want to cross-correlate) in the second time history. Let us
construct the statistic:
\begin{equation}
R^2(\tau \equiv k \Delta) = \sum_{i=i_{start}}^{i=i_{start}+N}
\frac{(n_{2,i} - s
n_{1,i+k})^2}{(\sigma_{2,i}^2+s^2\sigma_{1,i+k}^2)}
\end{equation}
Where $s$ is the scaling factor estimated as a ratio of the burst
counts detected by the instruments $s = \sum_i n_{1,i}/ \sum_j
n_{2,j}$. In the perfect case of identical detectors, with identical
energy ranges and arrival angles, and Poisson statistics, $R^2$ is
distributed as $\chi^2$ with $N$ degrees of freedom (dof). In
practice, there are several complicating factors. The detectors have
different energy ranges, different responses, and operate in
different background environments. Fortunately, for short GRBs some
of these complicating factors are less important: a) the background
variation on short time scales is small, and b) different spectral
evolutions which produce a significant lag between light curves
measured in different energy bands, is almost absent in short GRBs
\citep[e.g.,][]{norris01}.

\begin{figure}
\centering
\includegraphics[width=0.8\textwidth]{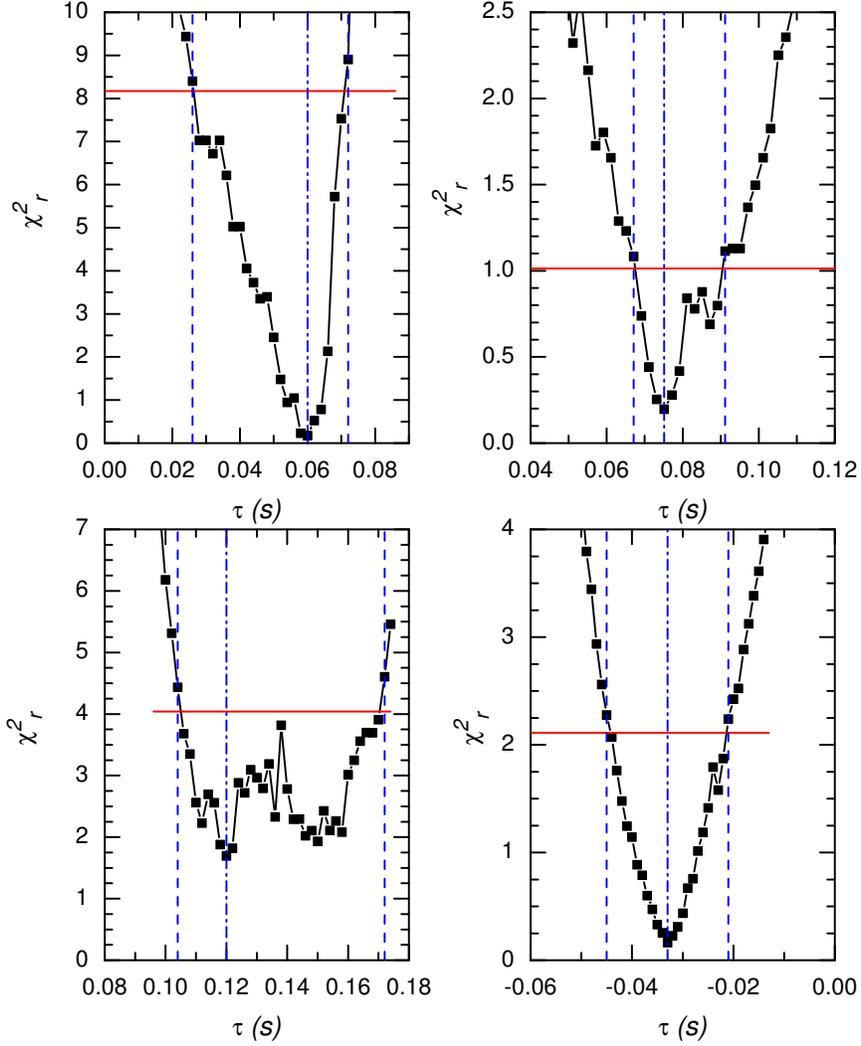}
\caption{Examples of cross-correlation curves $\chi^2_r(\tau)$.
Horizontal red lines denote 3$\sigma$ levels. Vertical blue lines
show the best cross-correlation time lag $\tau$ (dashed-dotted line)
and its 3$\sigma$ confidence interval (dashed lines). \textit{Top
Left:} GRB19971118\_T29008. Cross-correlation of the \textit{KW}
2~ms light curve with the BATSE 64 ms light curve; $\tau =
0.060(-0.034,+0.012)$~s (dof=2). \textit{Top Right:}
GRB20070321\_T67937. Cross-correlation of the \textit{KW} 2~ms light
curve with the WAM 1/64~s light curve; $\tau =
0.075(-0.008,+0.016)$~s (dof=12). \textit{Bottom Left:}
GRB20090715\_T62736. Cross-correlation of the \textit{KW} 2~ms light
curve with the SPI-ACS 50~ms light curve; $\tau =
0.120(-0.016,+0.052)$~s (dof=7). \textit{Bottom Right:}
GRB20100206\_T48606. Cross-correlation of the GBM 1~ms light curve
with the \textit{KW} 16~ms light curve; $\tau = -0.033 \pm 0.012$~s
(dof=9).}
\label{Fig_R2examples}
\end{figure}

To account for all deviations from the perfect case we adopt the
following approach: for a given $N$ (the number of bins used to
construct $R^2$) we calculate $\chi^2(N)$ corresponding to the
3$\sigma$ confidence level (that is, $\chi^2$ for which the
chi-square probability function $Q(\chi^2 | N) = 2.7 \times
10^{-3}$), and adopt the corresponding 3$\sigma$ level for the
reduced $R^2_r (\equiv R^2/N)$ of
\begin{equation}
\label{EqR2sigma3}
 R^2_{r, 3\sigma} = \chi^2_{r, 3\sigma}(N) + R^2_{r,min} - 1
\end{equation}
where $R^2_{r,min}$ is the minimum of $R^2_{r}(\tau)$; 1 is
subtracted, since $R^2_{r,min} \sim 1$ for the perfect case (in
practice it is often $>1$ and hence, $R^2_{r,3\sigma} > \chi^2_{r,
3\sigma}(N)$). To identify the 3$\sigma$ confidence interval for
$\tau$, we use the nearest points of the $R^2_r(\tau)$ curve above
the 3$\sigma$ level given by Eq.~(\ref{EqR2sigma3}) -– see the
examples in Figure~\ref{Fig_R2examples}. After the time lag $\tau$
and its errors $d_\pm(\tau)$ have been found, the time delay and its
uncertainty can be calculated as $\delta T = t_{02} - t_{01} +
\tau$; $d_\pm(\delta T)= d_\pm(\tau)$ (here we assume there is no
error in the absolute times $t_{01}$ and $t_{02}$). For simplicity
we will further refer to $R^2$ as $\chi^2$.

\section{LOCALIZATIONS: TRIANGULATION ANNULI}
Using the above methodology one or more triangulation annuli have
been obtained for 271 Konus-\textit{Wind} short bursts. Specific
details on the time delay determination for different pairs of
instruments are given in the subsections below.

\subsection{Annuli involving distant s/c}
Distant (interplanetary) spacecraft play an important role in GRB
triangulation.  Their long baselines make it possible to derive very
small error boxes for many bursts.  However, the detectors aboard
these missions tend to be smaller than ones in orbits closer to
Earth, and in some cases, are not dedicated GRB detectors, but
rather, are planetary experiments which have GRB detection modes.
Thus, they may have coarser time resolution and less sensitivity.
Also, the spacecraft clocks on these missions are not always
calibrated to UTC as accurately as the ones on missions closer to
Earth (or their calibration cannot be determined as accurately).  In
the present catalog, the data of four interplanetary missions have
been used: \textit{Ulysses}, \textit{NEAR}, \textit{Mars Odyssey},
and \textit{MESSENGER}. Of the four, only \textit{Ulysses} had a
dedicated GRB experiment.  The time resolutions of the four ranged
from 32 ms (\textit{Ulysses}, triggered mode) to 1 s
(\textit{MESSENGER}, \textit{NEAR}). When a short GRB is detected by
an experiment with time resolution much greater than the burst
duration, the result is usually an increase in the count rate in a
single time bin, which means that the timing uncertainty is
approximately one-half of the larger time resolution. The accuracy
of the spacecraft clocks have been verified in two ways. For
\textit{Ulysses}, commands were sent to the GRB experiment at
accurately known times, and, taking light-travel time and delays
aboard the spacecraft into account, the timing could be verified to
between several milliseconds and 125 milliseconds. (Although we
believe that the timing was accurate to several milliseconds,
technical issues often prevented us from verifying it.) In addition,
the timing of all the interplanetary missions can be verified using
triangulation of transient sources whose positions are well known by
other means: soft gamma-ray repeaters (SGRs) are one possibility,
and GRBs localized by the \textit{Swift} XRT or UVOT, for example,
are another.  For the purposes of this catalog, we have taken the
extremely conservative approach that no 3$\sigma$ cross-correlation
uncertainty is less than 125 ms.

In total 132 Konus short bursts were observed by distant s/c: 30 by
two distant s/c and 102 by one distant s/c. Among them 9 were
precisely localized by instruments with imaging capabilities. We do
not give here distant s/c annuli for these 9 bursts since they do
not improve the precise localizations.

As a result 150 annuli have been obtained using the distant s/c data
(two distant annuli for 27 bursts and one distant annulus for 96
bursts). Some of them have  already been presented in the IPN
catalogs: for BATSE bursts in~\citet{hurley99a,hurley99b,hurley11a},
 for \textit{BeppoSAX} bursts in~\citet{hurley10a}, and for \textit{HETE-2} bursts in~\citet{hurley11b}.

\begin{figure}
\centering
\includegraphics[width=0.7\textwidth]{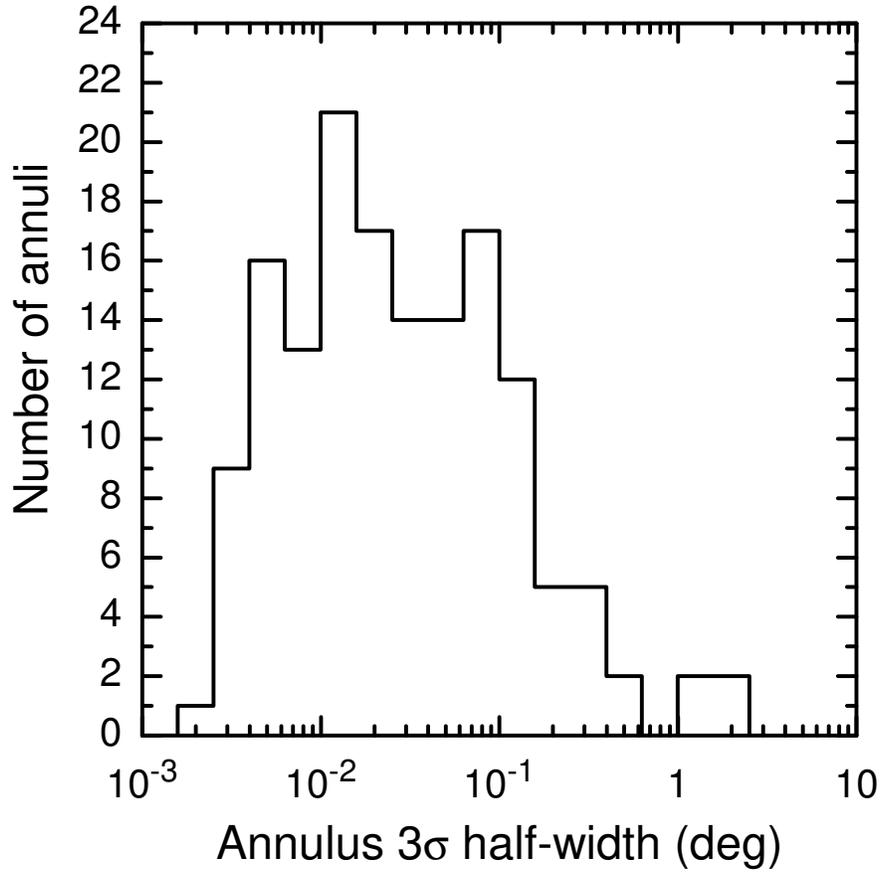}
\caption{Distribution of 3$\sigma$ half-widths (HWs) of the 150
triangulation annuli obtained using the distant s/c data. The
smallest HW is 0$\fdg$0024 (0.14\arcmin), the largest is 2$\fdg$21,
the mean is 0$\fdg$099 (5.9\arcmin), and the geometrical mean is
0$\fdg$028 (1.7\arcmin).}
\label{Fig_DistAnnuli}
\end{figure}

The histogram of Figure~\ref{Fig_DistAnnuli} shows the distribution
of 3$\sigma$ half-widths (HWs) of these 150 annuli. The smallest HW
is 0$\fdg$0024 (0.14\arcmin), the largest is 2$\fdg$21, the mean is
0$\fdg$099 (5.9\arcmin), and the geometrical mean is 0$\fdg$028
(1.7\arcmin).

\subsection{Annuli involving Konus-\textit{Wind}, \textit{INTEGRAL}, and a near-Earth s/c}
\begin{figure}
\centering
\includegraphics[width=0.7\textwidth]{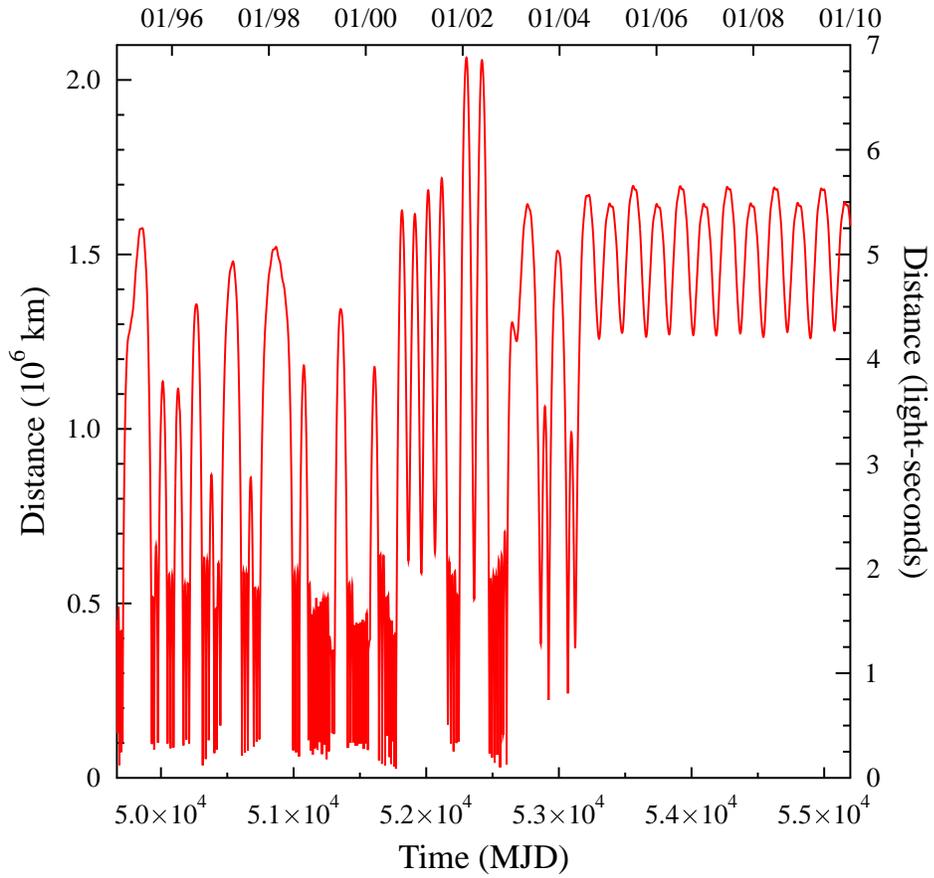}
\caption{\textit{Wind} distance from Earth as a function of time.
The maximum distance was $\simeq$7 lt-s in January and May 2002,
when it was in a Distant Prograde Orbit (DPO). Since 2004
\textit{Wind} has been in a Lissajous orbit at the L$_1$ libration
point of the Sun-Earth system at a distance of $\simeq$5 lt-s.}
\label{Fig_WindDistance}
\end{figure}

The Konus-\textit{Wind} (hereafter \textit{KW}) experiment plays a
special role in the IPN thanks to its unique set of characteristics:
continuous coverage of the full sky by two omnidirectional
spectrometers, orbit in interplanetary space that provides an
exceptionally stable background, wide energy range (10 keV -- 10 MeV
nominal; $\sim$20 keV -- 15 MeV at the present time), and a rather
high sensitivity of about $10^{-7}$~erg~cm$^{-2}$.  The \textit{KW}
duty cycle, defined as the time for data recovered divided by the
total operational time of the experiment, is about 95\%. It has
observed most of the IPN events, providing an important vertex in
the IPN at a distance of $\simeq$1--7 lt-s (see
Figure~\ref{Fig_WindDistance}).

In the triggered mode \textit{KW} records a burst time history in
three energy ranges G1, G2, G3 with nominal bounds 10--50 keV,
50--200 keV, and 200--750 keV, with a variable time resolution from
2 ms up to 256 ms \citep[for more details see][]{aptekar95}. The
time interval with the finest time resolution of 2 ms runs from
T$_0$-0.512 s to T$_0$+0.512~s (T$_0$ is the trigger time) and in
most cases covers the whole short burst, or at least its most
intense pulse, thereby allowing very accurate cross-correlation with
light curves of other instruments.

The \textit{KW} clock is accurate to better than 1~ms, and this
accuracy has been extensively verified by triangulation of many SGR
bursts and GRBs.

The best results for cross-correlation with \textit{KW} (i.e.,
minimum uncertainties in derived time delays) are provided by
near-Earth s/c with high effective areas, namely \textit{CGRO}
BATSE, \textit{BeppoSAX} GRBM, \textit{INTEGRAL}-SPI-ACS,
\textit{Suzaku}-WAM, \textit{Swift}-BAT, and \textit{Fermi}-GBM.

At present, triangulation with \textit{Fermi}-GBM usually provides
the best result (the narrowest annulus) thanks to the similar
designs of the \textit{KW} and GBM detectors (NaI(Tl)
scintillators), the high effective area of the GBM (several hundred
cm$^2$ for the combination of several NaI(Tl) detectors), and photon
time-tagging in 128 energy channels, that allows GBM light curves to
be obtained in the same energy ranges as \textit{KW} light curves
with any desired time binning.

Since the clocks of most of the near-Earth instruments are very
accurate, high count statistics combined with fine time resolution
often results in an uncertainty in time delays as low as several
milliseconds. Therefore, despite the rather small distance between
\textit{KW} and near-Earth s/c of several lt-s, the resulting
relative error in time delay, $d_{\pm}(\delta T)/D$, which
determines the width of the annulus (see
Eq.~(\ref{EqHW},\ref{EqHWsimple})) can be comparable to or sometimes
even smaller than for annuli involving distant s/c. Such small
uncertainties of several ms, and hence narrow annuli, can often be
derived for short bursts with sharp peaks or very fast rise and/or
decay times. On the other hand, bursts with smooth single-pulse
light curves usually give rather large cross-correlation
uncertainties in time delays, and hence, rather wide triangulation
annuli.

For \textit{Wind}, \textit{INTEGRAL}, and near-Earth s/c, ephemeris
uncertainties are negligible compared to uncertainties in time
delays and we do not take them in account.

\begin{figure}
\centering
\includegraphics[width=0.7\textwidth]{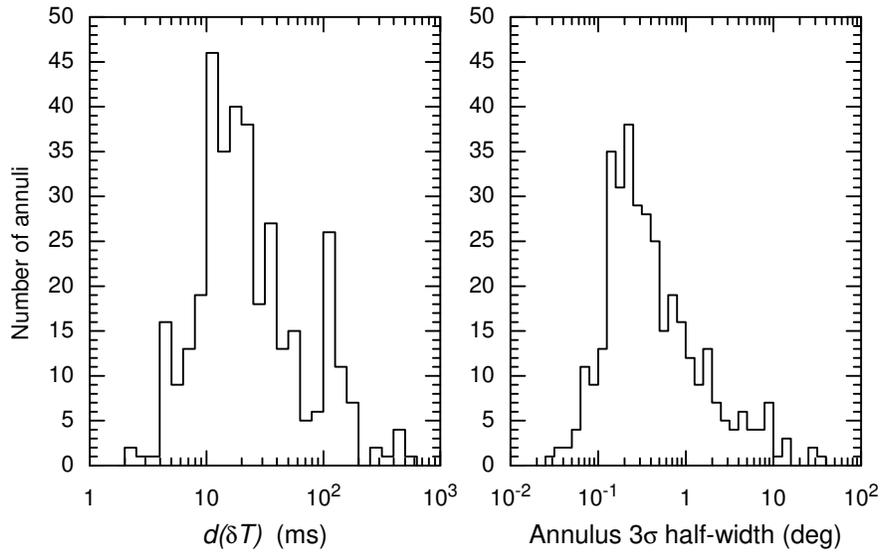}
\caption{Distributions of uncertainties in time delay $d(\delta T)
\equiv (d_{+}(\delta T) + |d_{-}(\delta T)|)/2$ and 3$\sigma$
half-widths (HWs) of the 356 triangulation annuli obtained using the
Konus-\textit{Wind} and near-Earth (or INTEGRAL) s/c data. The
smallest $d(\delta T)$ is 2~ms, the largest is 504~ms, the mean is
43~ms, and the geometrical mean is 23~ms. The smallest HW is
0\fdg027 (1.6\arcmin), the largest is 32\fdg2, the mean is 1\fdg30,
and the geometrical mean is 0\fdg43.}
\label{Fig_KWAnnuli}
\end{figure}

In total 356 \textit{KW}-near-Earth s/c and \textit{KW-INTEGRAL}
annuli have been obtained. The histograms of
Figure~\ref{Fig_KWAnnuli} show the distributions of uncertainties in
time delays and 3$\sigma$ half-widths of these annuli. The smallest
time delay uncertainty is 2~ms, the largest is 504~ms, the mean is
43~ms, and the geometrical mean is 23~ms. The smallest 3$\sigma$
half-width is 0\fdg027 (1.6\arcmin), the largest is 32\fdg2, the
mean is 1\fdg30, and the geometrical mean is 0\fdg43.

In the following subsections some details on triangulation involving
\textit{KW}, \textit{INTEGRAL}, and the near-Earth s/c are given.

\subsubsection{\textit{KW}-\textit{CGRO} (BATSE) triangulations\label{Sec_KWBATSEann}}
The Burst and Transient Source Experiment (BATSE) was a high-energy
astrophysics experiment  aboard the \textit{ Compton Gamma Ray
Observatory (CGRO)} \citep{fishman92}. Its Large Area Detectors
(LADs) measured burst time histories in four energy channels Ch1,
Ch2, Ch3, Ch4, with approximate channel boundaries:  25-55 keV,
55-110 keV, 110-320 keV, and $>$320 keV. The clock on the
\textit{CGRO} is accurate to 100 $\mu$s, and this accuracy was
verified through pulsar timing. Onboard software increases the
uncertainty in the BATSE trigger times to $\simeq$1~ms.

BATSE observed 52 \textit{KW} short GRBs: 44 in the triggered mode,
and 8 in the real-time mode. We derived \textit{KW}-BATSE annuli for
the 44 \textit{KW} short bursts observed in the triggered mode and
for 6 bursts observed by BATSE in real-time mode (these bursts were
observed by only \textit{KW} and BATSE).

For cross-correlation with triggered BATSE bursts we utilized
\textit{KW} light curves in the G2+G3 or in the G2 band with 2 or
16-ms resolution and BATSE concatenated light curves (DISCLA, PREB,
and DISCSC data types) in the Ch2+Ch3+Ch4 or in the Ch2+Ch3 band
with  a time resolution of 64 ms. For several GRBs such light curves
are not available and we utilized other types of BATSE data. Usually
we tried different energy bands to check the consistency of the
derived time delay and finally chose those with minimum $\chi^2$.
The cross-correlation curves for different bands may be slightly
shifted relative each other (by several ms) but the 3$\sigma$
intervals for cross-correlation lag $\tau$ are always in good
agreement.

The resulting $\chi^2_{r,min}$ range from 0.06 to 4.51 with a mean
of 0.81. The maximum $\chi^2_{r,min}$ of 4.51 (dof=6) is a clear
outlier in the distribution of bursts over $\chi^2_{r, min}$. It
corresponds to the exceptionally intense GRB19970704\_T04097 (BATSE
\#6293) with a peak count rate of $1.8 \times 10^5$ counts~s$^{-1}$
(\textit{KW} 2-ms time scale) and $6.9 \times 10^5$ counts~s$^{-1}$
(BATSE 64-ms time scale). Both light curves (\textit{KW} and BATSE)
must be significantly distorted due to dead-time and pile-up
effects. The derived statistical uncertainty in the time delay is
only 3~ms, so we added 6~ms systematic uncertainty to account for
the distortions.

The derived uncertainties in the time delays range from 5~ms to
84~ms with a mean of 24~ms, and a geometrical mean of 18~ms. The
resulting annuli 3$\sigma$ half-widths range from 0\fdg082 to
11\fdg0 with a mean of 1\fdg14, and a geometrical mean of 0\fdg60.
The widest annulus with half-width of 11\fdg0 was obtained for
GRB19991001\_T04950 (BATSE \#7781) -- at that time \textit{Wind} was
only 0.34 lt-s from Earth.

The distances between the center lines of the \textit{KW}-BATSE
annuli and the centers of BATSE locations range from 0\fdg007 to
7\fdg7 with a mean of 2\fdg23. For 14 bursts the BATSE error circle
does not intersect the \textit{KW}-BATSE annulus and the distances
from the closest boundary of the annulus range from 1.02$\sigma$ to
7.2$\sigma$
 \footnote{For GRB19961225\_T36436 (BATSE \#5725) the BATSE position given in the current BATSE catalog
 (R.A., Decl.(J2000), Err = 171\fdg31,  -2\fdg85, 1\fdg5) is 25\fdg2 away from the center line of the 1\fdg6
 wide \textit{KW}-BATSE annulus. This GRB is a ``mirror" case for BATSE, with a
 bimodal probability distribution for the location.
 The alternative solution is  R.A., Decl.(J2000) = 141\fdg4, -5\fdg5, with a statistical error of 1\fdg8
 (M. Briggs, private communication, 2011). This position is 7\fdg3 away
from the center line of the \textit{KW}-BATSE annulus. We use this
alternative BATSE position here.}.

\begin{figure}
\includegraphics[width=0.7\textwidth]{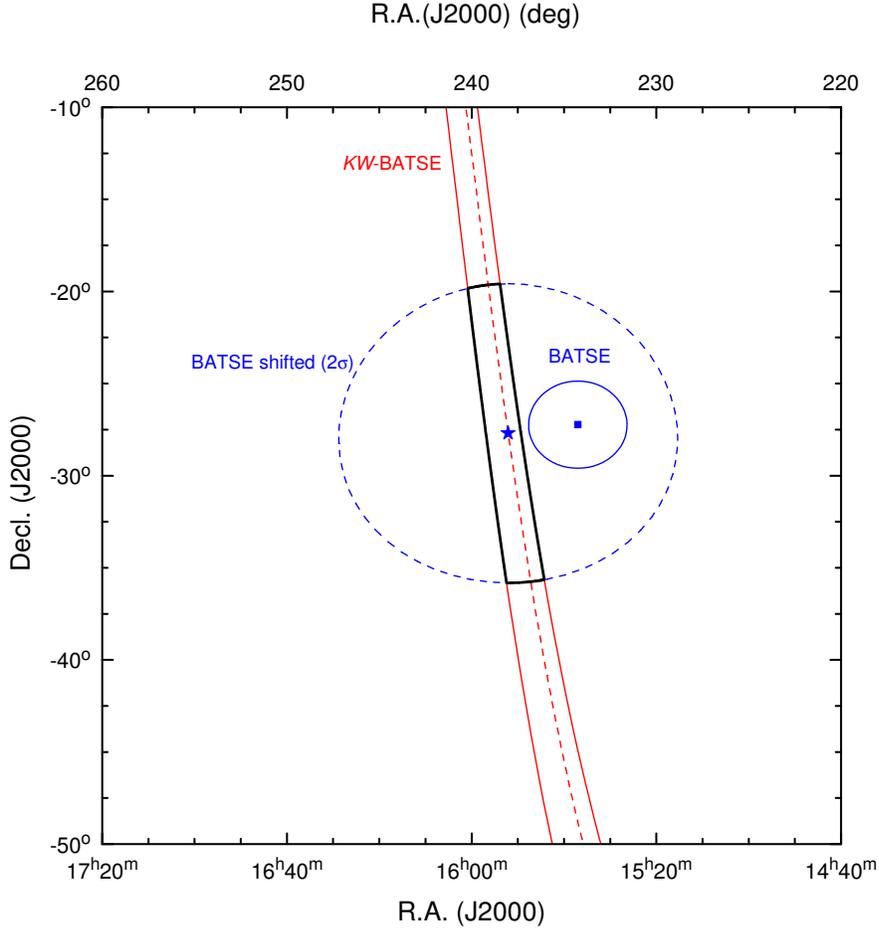}
\caption{IPN/BATSE localization of GRB19960420\_T16844 (BATSE
\#5439). The center of the BATSE error circle (R.A., Decl.(J2000),
Err = 234\fdg25, -27\fdg23, 2\fdg37) lies 3\fdg38 from the center
line of the 1\fdg67 wide \textit{KW}-BATSE annulus. The resulting
long box is shown by the solid black line and its center (that is,
the nearest point to the BATSE center at the annulus center line) is
indicated by the asterisk. The corners of the box are formed by the
intersection of the circle centered at the asterisk with a radius of
8$\fdg$12, that is the sum of the 2$\sigma$ BATSE error radius and
3\fdg38 systematics (dashed line), and  the \textit{KW}-BATSE
annulus.}
\label{Fig_KW_BATSE_loc}
\end{figure}

Of the 52 bursts, 16 were observed by only \textit{KW} and BATSE,
and 12 were observed by only \textit{KW}, BATSE, and
\textit{BeppoSAX}. For these bursts, we constrained the burst
location to a segment of the \textit{KW}-BATSE annulus using the
following method. We took the center of the segment to be the point
on the annulus center line nearest to the center of the BATSE
position, and then we derived the corners of the segment from the
intersection of the annulus and a circle centered at this point with
a radius equal to the sum of twice the BATSE 1$\sigma$ error and a
systematic error, taken to be the larger of 2\fdg0  and the distance
between the BATSE position and the center line of the annulus (a
core systematic error of $\simeq 2^\circ$ was found for BATSE
locations: see \citealt{briggs99}). This is illustrated in
Figure~\ref{Fig_KW_BATSE_loc}.

\subsubsection{\textit{KW}-\textit{Fermi} (GBM) triangulations}
The Gamma-Ray Burst Monitor (GBM) aboard the \textit{Fermi}
observatory is primarily designed to study of GRBs by making
observations in the $\sim$8 keV -- 40 MeV band \citep{meegan09}. The
GBM has the advantage of high effective area and time-tagged data.
The absolute timing of the GBM clock has an accuracy better than
20$\mu$s. The GBM TTE data contain counts in 128 energy channels
from $\sim$5~keV to 2~MeV, which enables the preparation of the GBM
light curve in three energy channels which are nearly the same as
those of  \textit{KW}.

GBM observed 34 \textit{KW} short GRBs. We derived \textit{KW}-GBM
triangulation annuli for all of them.

For cross-correlation we utilized \textit{KW} light curves in the
G2+G3 or in the G2 band with 2 or 16-ms resolution and GBM light
curves with 1 or 16 ms resolution made from the TTE data (only NaI
data were used).

The resulting $\chi^2_{r, min}$ range from 0.16 to 2.10 with a mean
of 0.90. The derived uncertainties in the time delays range from
2.5~ms to 136~ms with a mean of 22~ms, and a geometrical mean of
15~ms.  The resulting annuli 3$\sigma$ half-widths range from
0\fdg035 (2.1\arcmin) to 1\fdg65 with a mean of 0\fdg35, and a
geometrical mean of 0\fdg23.

\subsubsection{\textit{KW}-\textit{INTEGRAL} (SPI-ACS) triangulations}
The Anti-Coincidence Shield (ACS) of the SPI instrument on-board
\textit{INTEGRAL}, besides serving to veto the background in the
germanium spectrometer, is routinely used as a nearly
omnidirectional detector for gamma-ray bursts~\citep{kienlin03}. It
measures burst light curves with a time resolution of 50~ms in a
single energy range above $\sim$80~keV (for more details see
\citealt{lichti00}). A systematic error in the ACS timing of
125$\pm$10~ms has been found~\citep{rau04} and all SPI-ACS lc have
been corrected automatically for this error after 2004 April;
corrections for it were applied by hand to the data prior to this
date.

This systematic uncertainty is related to the approximate nature of
conversion from on-board time to UTC used in the producing SPI-ACS
light curves in real time (within just seconds after the event at
\url{ftp://isdcarc.unige.ch/arc/FTP/ibas/spiacs/}).

On the other hand, the time conversion used in archived and near
real-time data is precise. Recently it was shown that the drift of
the ACS clock with respect to the germanium detector clock during
the whole \textit{INTEGRAL} mission is around 1 ms~\citep{zhang10},
thereby reducing the systematic uncertainty of ACS timing from 10 ms
to 1 ms.

SPI-ACS light curves corrected for systematic shifts and
characterized by high timing precision (at least down to 1 ms) are
available in the \textit{INTEGRAL} data archive since revision 3.
The archived and equally precise near real-time data (available
within hours after the observation) are accessible through web
services \url{http://isdc.unige.ch/\~{}savchenk/spiacs-online/} and
\url{http://www.isdc.unige.ch/heavens/}. They are routinely used for
near real-time triangulation.

SPI-ACS observed 139 \textit{KW} short GRBs. We derived
\textit{KW}-SPI-ACS annuli for 103 of them.

For cross-correlation we utilized \textit{KW} light curves in the
G2+G3 or in the G3 band with 2 or 16-ms resolution.

The resulting $\chi^2_{r,min}$ range from 0.04 to 3.96 with a mean
of 1.02. The derived uncertainties in the time delays range from
4~ms to 175~ms with a mean of 24~ms, and a geometrical mean of
19~ms. The resulting annuli 3$\sigma$ half-widths range from
0\fdg047 (2.8\arcmin) to 4\fdg3 with a mean of 0\fdg41, and a
geometrical mean of 0\fdg29.

\subsubsection{\textit{KW}-\textit{Suzaku} (WAM) triangulations}
The Wide-band All-sky Monitor (WAM) is the active shield of the Hard
X-ray detector aboard the \textit{Suzaku} mission \citep{yamaoka09}.
In the triggered mode it measures light curves with a time
resolution of 1/64~s in four energy channels which cover the
$\simeq$50--5000 keV range. In the real-time mode the time
resolution is 1~s.

It was established that the \textit{Suzaku}-WAM timing is consistent
with negligible systematic uncertainties~\citep{yamaoka09}.

WAM observed 61 \textit{KW} short GRBs: 51 in the triggered mode and
10 in the real-time mode. We derived \textit{KW}-WAM annuli for 45
triggered bursts.

For cross-correlation we utilized \textit{KW} light curves in the
G2+G3 or in the G3 band with 2 or 16-ms resolution and WAM light
curves summed over four energy channels of 1 to 4 WAM detectors with
the strongest response.

The resulting $\chi^2_{r,min}$ range from 0.21 to 1.78 with a mean
of 1.03. The derived uncertainties in the time delays range from
4~ms to 104~ms with a mean of 20~ms, and a geometrical mean of
14~ms. The resulting annuli 3$\sigma$ half-widths range from
0\fdg060 (3.6\arcmin) to 2\fdg44 with a mean of 0\fdg30, and a
geometrical mean of 0\fdg21.

\subsubsection{\textit{KW-BeppoSAX} (GRBM) triangulations}
The \textit{BeppoSAX} Gamma-Ray Burst Monitor (GRBM) was the
anticoincidence shield of the high energy experiment PDS
\citep[Phoswich Detection System;][]{frontera97}.

In the triggered mode it measured light curves with a time
resolution of 7.8125~ms in the 40--700~keV range; in the real-time
mode the time-resolution was 1~s \citep[for more details
see][]{frontera09}.

GRBM observed 50 \textit{KW} short GRBs: 41 in the triggered mode
and 9 in the real-time mode. We derived \textit{KW}-GRBM annuli for
38 bursts observed in the triggered mode and for one burst observed
in the real-time mode (this burst was observed by only \textit{KW}
and GRBM).

For cross-correlation we utilized \textit{KW} light curves in the G2
or in the G2+G3 band with 2 or 16-ms resolution and GRBM light
curves rebinned to 32 ms.

The resulting $\chi^2_{r,min}$ range from 0.25 to 12.1 with a mean
of 1.43. The maximum $\chi^2_{r,min}$ of 12.1 (dof=6) is a clear
outlier in the distribution of bursts over $\chi^2_{r, min}$. It
corresponds to the exceptionally intense GRB19970704\_T04097 with a
peak count rate of 1.8$\times 10^5$ counts~s$^{-1}$ (\textit{KW}
2-ms time scale) and 1.5$\times 10^5$ counts~s$^{-1}$ (GRBM 32-ms
time scale). Both light curves (\textit{KW} and GRBM) must be
significantly distorted due to dead-time and pile-up effects. The
derived statistical uncertainty in the time delay is only 2~ms, so
we increased this uncertainty to 6~ms to account for the
distortions.

The derived statistical uncertainties in the time delays range from
4.5~ms to 216~ms with a mean of 32~ms, and a geometrical mean of
18~ms.

Comparison of the initially derived annuli with other available IPN
annuli as well as comparison of the GRBM and BATSE light curves for
common bursts has revealed a systematic error in the GRBM timing up
to 100 ms. Since this error varies from burst to burst (both in
value and in sign), we had to introduce 100 ms systematic error for
\textit{KW}-SAX triangulations. This leads to a significant
broadening of the annuli, so their final 3$\sigma$ half-widths range
from 1\fdg23 to 32\fdg2 with a mean of 5\fdg30, and a geometrical
mean of 3\fdg87.

\subsubsection{\textit{KW}-\textit{Swift} (BAT) triangulations}
The \textit{Swift} Burst Alert Telescope (BAT) is a highly
sensitive, large field of view (FOV) coded aperture telescope that
detects and localizes GRBs in real time \citep{barthelmy05}. When a
burst occurs outside its FoV, it cannot be imaged, but the BAT light
curve can be used for triangulation. For such bursts the 64 ms light
curves in the four standard BAT energy channels (15--25 keV, 25--50
keV,  50--100 keV, and 100--350 keV) are always available. For some
bursts, TTE data are available, enabling any desired energy and time
binning.

BAT observed 44 \textit{KW} short bursts outside its FoV. We derived
\textit{KW}-BAT annuli for 23 of them.

For cross-correlation we utilized \textit{KW} light curves in the G2
or in the G2+G3 band with 2 or 16-ms resolution and BAT 64 ms light
curves usually taken in the energy range above 50 keV (which often
provides the best S/N and corresponds better to the \textit{KW}
energy band).

The resulting $\chi^2_{r,min}$ range from 0.25 to 7.48 with a mean
of 1.41. The maximum $\chi^2_{r,min}$ of 7.48 (dof=6) is a clear
outlier in the distribution of bursts over $\chi^2_{r, min}$. It
corresponds to the exceptionally intense GRB20060306\_T55358 with
strong spectral evolution and a peak count rate of $1.9 \times 10^5$
counts~s$^{-1}$ (\textit{KW} 2-ms time scale). The derived
statistical uncertainty in the time delay is only 5~ms, so we added
10~ms systematic  uncertainty.

The derived uncertainties in the time delays range from 5~ms to
64~ms with a mean of 22~ms, and a geometrical mean of 18~ms. The
resulting annuli 3$\sigma$ half-widths range from 0\fdg059
(3.5\arcmin) to 1\fdg18 with a mean of 0\fdg41, and a geometrical
mean of 0\fdg29.

\subsubsection{\textit{KW}-\textit{Coronas-F} (Helicon) triangulations}
The Helicon gamma-ray spectrometer was one of the instruments
onboard the \textit{Coronas-F} solar space
observatory~\citep{oraevskii02}. It was similar to the \textit{KW}
spectrometer in the characteristics of its two detectors and in the
data presentation structure. The similar design of both instruments
enabled good cross-correlations of burst light curves.

Helicon observed 14 \textit{KW} short GRBs. We derived
\textit{KW}-Helicon annuli for all of them.

The resulting $\chi^2_{r,min}$ range from 0.25 to 2.67 with a mean
of 1.02. The derived uncertainties in the time delays range from
4~ms to 80~ms with a mean of 25~ms, and a geometrical mean of 17~ms.
The resulting annuli 3$\sigma$ half-widths range from 0\fdg045
(2.7\arcmin) to 1\fdg15 with a mean of 0\fdg38, and a geometrical
mean of 0\fdg25.

\subsubsection{\textit{KW}-\textit{Cosmos} (Konus-A,A2,A3) triangulations}
Konus-A, Konus-A2, and Konus-A3 were gamma-ray spectrometers aboard
the \textit{Cosmos} spacecraft 2326, 2367, and 2421, respectively. A
brief description of the Konus-A instrument is given in
\citet{aptekar98}. Konus-A2 and Konus-A3 were similar to Konus-A in
the characteristics of its detectors and in the data presentation
structure.

They observed 5 \textit{KW} short GRBs in the triggered mode. We
derived \textit{KW}-\textit{KA} annuli for 4 of them.

The resulting $\chi^2_{r,min}$ range from 0.73 to 1.42 with a mean
of 1.07. The derived uncertainties in the time delays range from
4~ms to 56~ms with a mean of 35~ms. The resulting annuli 3$\sigma$
half-widths range from 0\fdg15 to 1\fdg20 with a mean of 0\fdg79.

\subsubsection{\textit{KW-RHESSI} triangulations}
The \textit{Reuven Ramaty High-Energy Solar Spectroscopic Imager}
(\textit{RHESSI}) is a high resolution spectrometer designed to
study high-energy emission from solar flares over a broad energy
range from 3 keV to 17 MeV \citep{lin02,smith02}. The data are
collected in the TTE mode enabling arbitrary energy and time
binning.

\textit{RHESSI} observed 58 \textit{KW} short GRBs. We derived
\textit{KW}-RHESSI annuli for 32 bursts.

For cross-correlation we utilized \textit{KW} light curve in the G2,
in the G1+G2, or in the G2+G3 band with 2, 16, 64, and 256-ms
resolution (depending on burst intensity).

The resulting $\chi^2_{r,min}$ range from 0.36 to 2.62 with a mean
of 1.07. The derived uncertainties in the time delays range from
2~ms to 184~ms with a mean of 36~ms, and a geometrical mean of
20~ms. The resulting annuli 3$\sigma$ half-widths range from
0\fdg027 (1.6\arcmin) to 2\fdg71 with a mean of 0\fdg53, and a
geometrical mean of 0\fdg30.

\subsubsection{\textit{KW-HETE-2} (FREGATE) triangulations}
The gamma-ray detector of \textit{HETE-2}, called FREGATE, was
designed to detect gamma-ray bursts in the energy range 8–-400 keV
\citep{ricker03,atteia03}.

In the triggered mode it measures light curves with a time
resolution of 1/32~s in the 8--400~keV range; in the real-time mode
the time-resolution is 0.1638~s.

FREGATE observed 16 \textit{KW} short GRBs: 8 in the triggered mode
and 8 in the real-time mode. In most cases the FREGATE response is
significantly weaker than the responses of other instruments flying
on low-Earth s/c, so we used the FREGATE data only for a few cases
when no other low-Earth s/c detected the burst. We derived
\textit{KW-HETE} annuli for 4 bursts observed in the triggered mode.

For cross-correlation we utilized \textit{KW} light curve in the G2
or in the G2+G3 band with 2 or 16-ms resolution.

The resulting $\chi^2_{r,min}$ range from 0.50 to 1.43 with a mean
of 0.96. The derived uncertainties in the time delays range from
56~ms to 168~ms with a mean of 102~ms.  The resulting annuli
3$\sigma$ half-widths range from 0\fdg95 to 1\fdg47 with a mean of
1\fdg10.

\subsubsection{\textit{KW-AGILE} (MCAL) triangulations}
The mini-calorimeter (MCAL) aboard the \textit{AGILE} mission is a
spectrometer sensitive to gamma-rays in the energy band $\simeq$0.35
-- 100 MeV \citep{tavani09}. MCAL has the advantage of time-tagged
data.

MCAL observed 24 \textit{KW} short GRBs: 22 in the triggered mode
and 2 in the real-time mode. In many cases the MCAL response is
rather weak due to its high energy threshold and strong attenuation
by the GRID instrument, so we used the MCAL data only for several
intense bursts. In total we derived 9 \textit{KW}-MCAL annuli.

For cross-correlation we utilized \textit{KW} light curve in the G3
or in the G2+G3 band with 2 or 16-ms resolution.

The resulting $\chi^2_{r,min}$ range from 0.29 to 2.26 with a mean
of 1.08. The derived uncertainties in the time delays range from
5~ms to 21~ms with a mean of 13~ms.  The resulting annuli 3$\sigma$
half-widths range from 0\fdg071 (4.3\arcmin) to 0\fdg60 with a mean
of 0\fdg21.

\subsubsection{INTEGRAL -- near-Earth s/c triangulations}
Even without the planetary missions, the mini-network of low-Earth
orbiters, plus \textit{INTEGRAL} and Konus-\textit{Wind}, often make
it possible to obtain error boxes for many bursts. Since
\textit{INTEGRAL} orbits at distances $\lesssim$0.5 lt-s, which are
much smaller than the \textit{Wind}-to-Earth distance $\simeq$5
lt-s, \textit{KW-INTEGRAL} and \textit{KW}-near-Earth s/c annuli
intersect at grazing incidence, resulting in one or two long boxes.
In some cases, the intersection of \textit{KW}-near-Earth s/c and
\textit{INTEGRAL}-near-Earth s/c results in smaller localization
regions.

In total 11 \textit{INTEGRAL}-near-Earth s/c annuli have been
obtained. The resulting annuli half-widths range from 1\fdg0 to
14\fdg0 with a mean of 5\fdg9.

\subsection{Verifying triangulation annuli\label{Sec_AnnVerification}}
Of the 271 Konus short bursts localized by IPN, 17 were precisely
localized by instruments with imaging capabilities: 15 by
\textit{Swift}-BAT (one of them, GRB 090510, was also localized by
\textit{Fermi}-LAT), 1 by \textit{HETE-2} (WXM and SXC), and 1 by
\textit{INTEGRAL} IBIS/ISGRI.

We  utilized these bursts to verify our triangulations. For these 17
bursts, 21 \textit{KW}-near-Earth s/c and 12
\textit{KW}-\textit{INTEGRAL} annuli were obtained (we have not used
the light curves of the instruments which imaged the burst, since
the instrument response for imaged bursts is different from those
detected outside the FoV and used for IPN triangulations). In each
case the triangulation annuli are in agreement with the known
position of the source, thereby confirming the reliability of our
triangulations. Indeed, such tests constitute ``end-to-end"
calibrations, as they confirm not only spacecraft timing and
ephemeris information, but also the cross-correlations of the
various time histories and derivations of the annuli.

\begin{figure}
\centering
\includegraphics[width=0.7\textwidth]{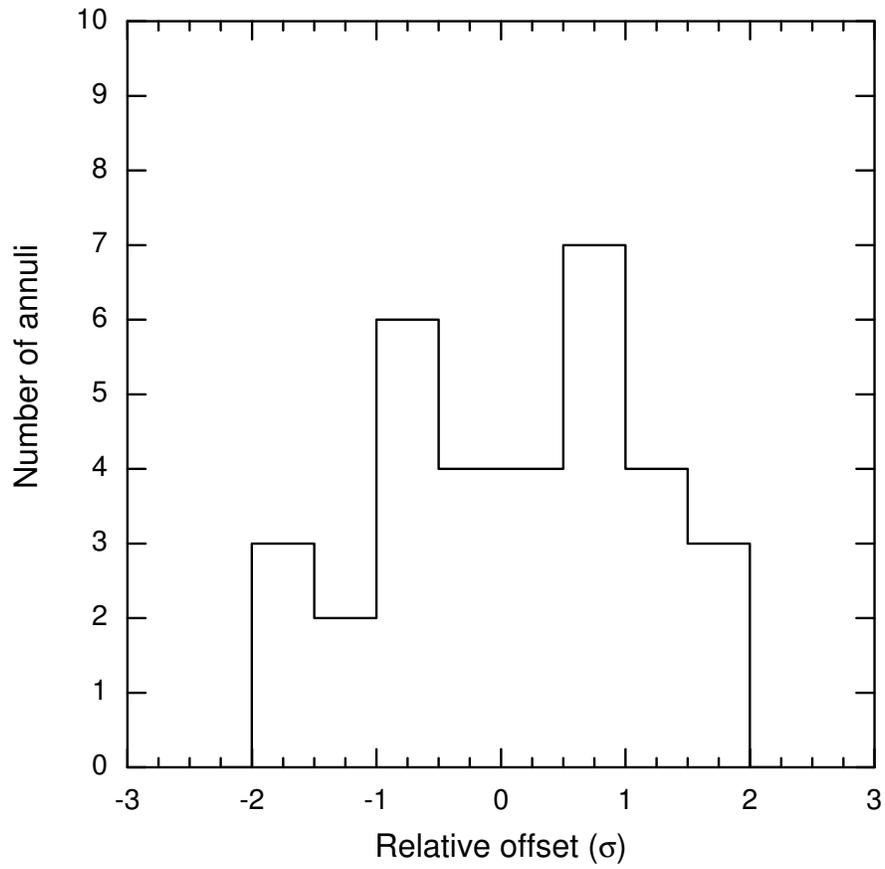}
\caption{Distribution of the offsets of the accurate GRB positions
from the center lines of the 33 \textit{KW}-near-Earth (or
\textit{INTEGRAL}) s/c annuli.}
\label{Fig_AnnuliVerification}
\end{figure}

The histograms of Figure~\ref{Fig_AnnuliVerification} show the
distribution of relative source offsets from the center lines of the
annuli. One can see that all offsets (in absolute values) are less
than 2$\sigma$. The minimum offset of the precise position is
-2.0$\sigma$, the maximum is 1.9$\sigma$, the average is
0.04$\sigma$, and the standard deviation is 1.1$\sigma$.

Besides this verification, the consistency of several
\textit{KW}-near-Earth s/c annuli often obtained for a given burst
with each other and with distant s/c annuli (when available)
confirms the reliability of our \textit{KW}-near-Earth
triangulations, many of which have time delay uncertainties less
than 10--20 ms (see Figure~\ref{Fig_KWAnnuli}).

\section{LOCALIZATIONS: ADDITIONAL CONSTRAINS}
In addition to triangulation annuli, several other types of
localization information are included in this catalog. They are
ecliptic latitude range, autonomous burst localizations obtained by
\textit{CGRO} BATSE, \textit{BeppoSAX} GRBM, and \textit{Fermi} GBM,
and Earth- or Mars-blocking (MESSENGER is in an eccentric orbit
around Mercury, so Mercury-blocking is quite rare). This additional
information helps constrain the triangulation position, i.e., to
choose one of two triangulation boxes, or to eliminate portions of a
single annulus.

\subsection{Ecliptic latitudes}
The ecliptic latitudes of the bursts are derived by comparing the
count rates of the two \textit{KW} detectors taken in the waiting
mode with 1.472 s or 2.944 s time resolution. The axis of the
detector S2 points towards the north ecliptic pole, and the axis of
 S1 points toward the south ecliptic pole. In addition to
statistical uncertainties, the ecliptic latitude determination is
subject to systematic uncertainties due to, among other things,
time-variable cosmic X-ray sources and absorption by other
instruments aboard the spin-stabilized \textit{Wind} spacecraft. The
estimated ecliptic latitudes can be taken to be at the 95\%
confidence level.

The ecliptic latitude range, namely the best estimate of $b$, and
the lower and upper limits $b_{min}$, $b_{max}$ can be considered to
be an annulus centered at the north or south ecliptic pole, with a
half-angle $\theta = 90^\circ-|b|$ and half-widths $d_{-}(\theta) =
b_{min}-b$, $d_+(\theta) = b_{max} - b$.

\subsection{Planet-blocking}

Planet-blocking is specified by the right ascension and declination
of the planet's center and its radius.  When a spacecraft in low
Earth or Mars orbit observes a burst, the planet blocks up to
$\approx$ 3.7 sr of the sky. The source position must be outside
this occulted part of the sky.

The allowed part of the sky can be described as a degenerate annulus
centered at the direction opposite to the planet's center, with a
half-angle $\theta=0$ whose widths $d_{-}(\theta)=0$, $d_{+}(\theta)
= \sin^{-1} (R_{planet}/R)$, where $R$ is the radius of the s/c
orbit (here we neglect the oblateness of the planet and absorption
in its atmosphere).

\subsection{Autonomous localizations}
A principle of autonomous burst localization using a system of
detectors posessing anisotropic angular sensitivity was suggested by
\citet{golenetskii74} and first implemented in the KONUS instruments
on the Venera 11 and 12 missions \citep{mazets81}.

Similar localization systems consisting of different numbers of
detectors have been placed on \textit{CGRO} (BATSE),
\textit{BeppoSAX} (GRBM), and \textit{Fermi} (GBM).  These
autonomous localizations, derived by comparing the count rates of
various detectors, are affected by Earth albedo and absorption by
spacecraft materials, among other things, and their shapes are in
general complex. The error circles are approximations to these
shapes. They are centered at the point which is the most likely
arrival direction for the burst, and their radii are defined so that
their areas are equal to the 1$\sigma$(BATSE, GBM) or 90\%
confidence (GRBM) statistical-only true error regions. All these
localizations also have systematic errors of several degrees or
more.

These error circles can also be described as degenerate annuli
centered at the most likely arrival direction for the burst, with a
half-angle $\theta=0$ whose widths $d_{-}(\theta)=0$, $d_{+}(\theta)
= r$, where $r$ is the positional error.

\section{LOCALIZATIONS: RESULTS}
Table~\ref{Table_Annuli} summarizes localization information for 271
Konus short bursts. The first column gives the burst designation
(see Table~\ref{Table_Basic}). The second column gives the number of
localization constraints (the number of rows with localization
information for the burst). The six subsequent columns give
localizations expressed as a set of annuli: the third column gives
the source of the location: either sc1--sc2 (triangulation annulus
derived using sc1 and sc2), or `Ecl.Band' (range of ecliptic
latitudes), or `Instr' (name of the instrument which autonomously
localized the bursts), or `Occ.sc' (planet-blocking); columns 4--8
list the right ascension and declination of the annulus center
(J2000), the annulus radius $\theta$, and the 3$\sigma$
uncertainties in the radius $d_{-}(\theta)$, $d_{+}(\theta)$.

Planet-blocking is given only if it constrains the location. The
ecliptic latitude range is given for all bursts. All available
autonomous localizations are given.

The \textit{Swift}-BAT localizations are taken from the second
\textit{Swift} BAT catalog covering 2004 December 19 to 2009
December 21 \citep{sakamoto11}, and for the latest bursts from the
GCN Circulars with BAT refined positions.

The \textit{HETE-2} localizations for GRB 040924
(=GRB20040924\_T42735) is taken from \citet{arimoto06}.

The IBIS/ISGRI localization for GRB 070707 (=GRB20070707\_T58122) is
taken from \citet{gotz07}.

The BATSE localizations are taken from the current catalog on the
BATSE
website~\footnote{\url{http://www.batse.msfc.nasa.gov/batse/grb/catalog/current/}},
as well as from the BATSE untriggered burst catalogs \citep{stern01,
kommers00}\footnote{Since the catalog by \citet{stern01} contains
the localizations for all 8 Konus short bursts detected by BATSE in
the real-time mode, and the catalog by \citet{kommers00} misses some
of them, the given localizations are solely from \citet{stern01}.}.

The \textit{BeppoSAX} localizations are taken either from the GRBM
catalog \citep{frontera09} or from the IAU and GCN Circulars.

The GBM localizations are taken from the first \textit{Fermi} GBM
catalog  covering 2008 July 12 to 2010 July 11 \citep{paciesas12},
the GCN Circulars, or from the latest version of the corresponding
`glg\_tcat*.fit' file in the GBM data
archive~\footnote{ftp://legacy.gsfc.nasa.gov/fermi/data/gbm/bursts}.

\subsection{Boxes}
For those bursts which were detected by three or more well separated
s/c, a triangulation box can be derived.

In general, the intersection of two annuli involving distant s/c
gives a small box with an area as small as 1 arcmin$^2$.

The intersection of two annuli derived from a distant s/c,
Konus-\textit{Wind}, and a near-Earth s/c usually gives an elongated
box, which nevertheless in most cases has a small area of several
hundred arcmin$^2$. In some cases the intersection of annuli derived
from a single distant s/c, Konus-\textit{Wind}, and a near-Earth s/c
can give a smaller error box than annuli derived using two distant
s/c.

Long boxes were derived for bursts not observed by any distant s/c,
but observed by \textit{KW}, \textit{INTEGRAL} SPI-ACS, and one or
more near-Earth s/c.
In such cases, the box is formed by a \textit{KW}--near-Earth s/c
annulus and an \textit{INTEGRAL}--near-Earth s/c annulus, or by a
\textit{KW}--near-Earth s/c annulus and a
\textit{KW}--\textit{INTEGRAL} annulus intersecting at grazing
incidence.

In all cases, if the three s/c which formed the box were nearly
aligned, the annuli intersect at grazing incidence, resulting in a
long box.

In total we derived 162 error boxes for Konus short bursts: 27 for
bursts observed by two distant s/c, 84 for bursts observed by one
distant s/c and at least one near-Earth s/c, and 51 for bursts
observed by only \textit{KW}, \textit{INTEGRAL}, and one or more
near-Earth s/c. In some cases these error regions are actually long
arcs rather than boxes (in particular this is a case when the burst
was not observed by a distant s/c), but for simplicity we still
refer to them as boxes since they are formed by intersection of two
or more triangulation annuli.

\subsection{Segments}
For those bursts which were detected only by \textit{KW} and another
s/c, or by \textit{KW} and one or more near-Earth s/c, the resulting
localization is formed by a triangulation annulus (the narrowest in
the case of several \textit{KW}-near-Earth s/c annuli) and
additional constraints. These localizations consist of the entire
annulus (in the case where it is entirely inside the allowed
ecliptic latitude band and there are no other constraints) or one or
two annulus segments, formed by the intersection of the annulus with
the ecliptic latitude band, and/or by exclusion of the occulted part
of the annulus, or by combination with the BATSE localization (see
section~\ref{Sec_KWBATSEann}).

114 bursts had this kind of localization: 20 were bursts observed by
\textit{KW} and a distant s/c (of these, 3 were also observed by a
near-Earth s/c in real-time mode, but \textit{KW}-near-Earth s/c
annuli have not been derived for them), and 94 were bursts observed
by only \textit{KW} and one or more near-Earth s/c.

\subsection{Resulting error regions\label{Sec_ErrorRegions}}
Table~\ref{Table_Boxes} gives the description of the final IPN error
regions for 254 Konus short bursts (this sample does not include the
17 imaged bursts). The nine columns contain the following
information: (1) the burst designation (see
Table~\ref{Table_Basic}), (2) the number of error regions for the
burst, $N_{r}$: 1 or 2, (3) the number of corners of the region,
$N_{c}$, (4) the region type: `B' (box), `LB' (long box: box with
maximum dimension $>10^\circ$), `S' (segment), or `A' (annulus), (5)
the area (for two regions the sum of their areas), (6) the maximum
dimension of the region (that is the maximum angular distance
between two points at the region boundary; for segments larger than
half an annulus, this is just the outer diameter of the annulus),
(7) the right ascension of the center of the error region, in the
first row, and the right ascensions of the corners in the following
$N_{c}$ rows, and (8) the declination of the center of the error
region, in the first row, and the declinations of the corners in the
following $N_{c}$ rows (if there are two error regions, additional
$N_{c}$+1 rows are given: the center of the second region and its
corners, so the total number of rows for such a burst is
2($N_{c}$+1)). All coordinates are J2000.

In general, a simple, four-corner error region description is
inaccurate and the curvature of the annuli should be taken into
account. Only in cases where the maximum dimension of the error
region is less than several degrees, can the box be reasonably well
represented by its four corners. In other cases, especially when the
region is a long arc or annulus segment, the given corners and
center are intended to roughly indicate the position of the region
on the sky. Figures showing the IPN localization (all derived annuli
and the resulting error region(s)) can be found at the Ioffe Web
site~\footnote{\url{http://www.ioffe.ru/LEA/ShortGRBs\_IPN/}}.

\begin{figure}
\centering
\includegraphics[width=0.7\textwidth]{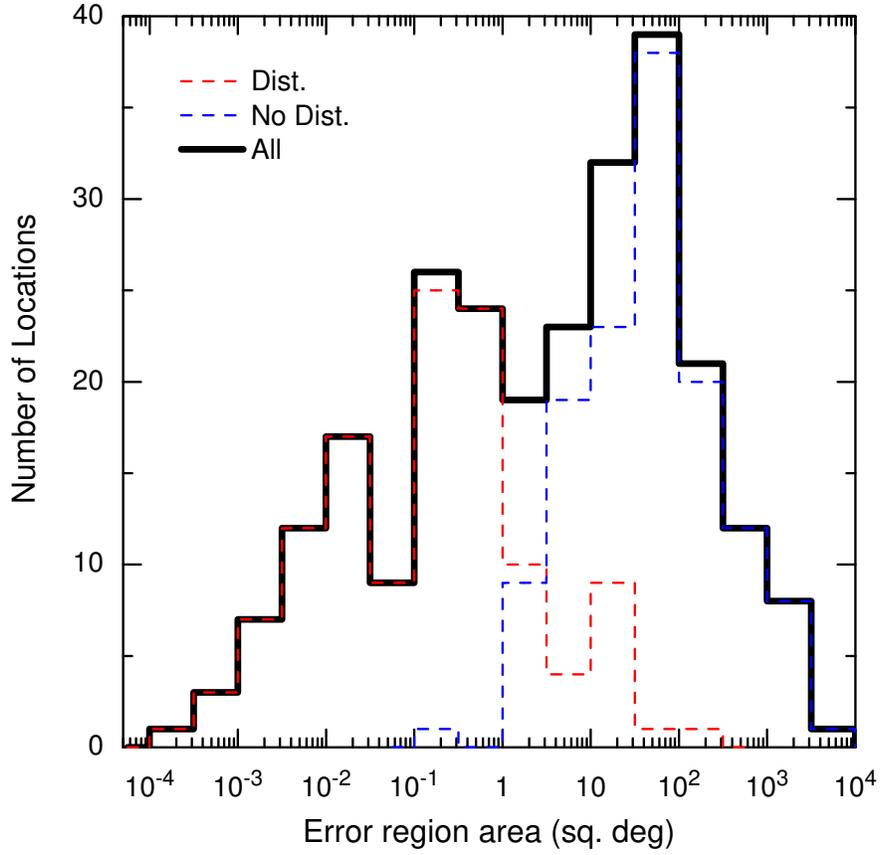}
\caption{Distributions of error region areas for 123 Konus short
bursts observed by at least one distant s/c (red dashed line), 131
bursts not observed by any distant s/c (blue dashed line), and all
254 bursts (17 imaged bursts are not counted).}
\label{Fig_AreaStat}
\end{figure}

A histogram of IPN error region areas is shown in
Figure~\ref{Fig_AreaStat}. For bursts observed by distant s/c the
areas range from 2.40$\times 10^{-4}$~sq.~deg (0.86 arcmin$^2$) to
142.1~sq.~deg with a mean of 3.49~sq.~deg, and a geometrical mean of
0.141~sq.~deg. For bursts without distant s/c detections the areas
range from 0.210~sq.~deg to 4420~sq.~deg with a mean of 242~sq.~deg,
and a geometrical mean of 46.2~sq.~deg.

\section{COMMENTS ON SPECIFIC EVENTS}
GRB~051103 (=GRB20051103\_T33943) may in fact be a giant SGR flare
in the nearby M81 group of interacting galaxies as was suggested by
\citet{frederiks07}. The final IPN localization of this event along
with further exploration of this possibility are given
in~\citet{hurley10b}. See also \cite{ofek06} for implications of the
optical and radio followup observations and \citet{abadie12} for
implications of the gravitational-wave non-detection of this event.

GRB~070201 (=GRB20070201\_T55390) is likely a giant SGR flare from
the Andromeda galaxy \citep{mazets08}. See also \citet{abbott08} for
implications of the gravitational-wave non-detection of this event,
and  \citet{ofek08} for implications of the optical afterglow and
X-ray periodic source non-detections.

\begin{figure}
\centering
\includegraphics[width=0.7\textwidth]{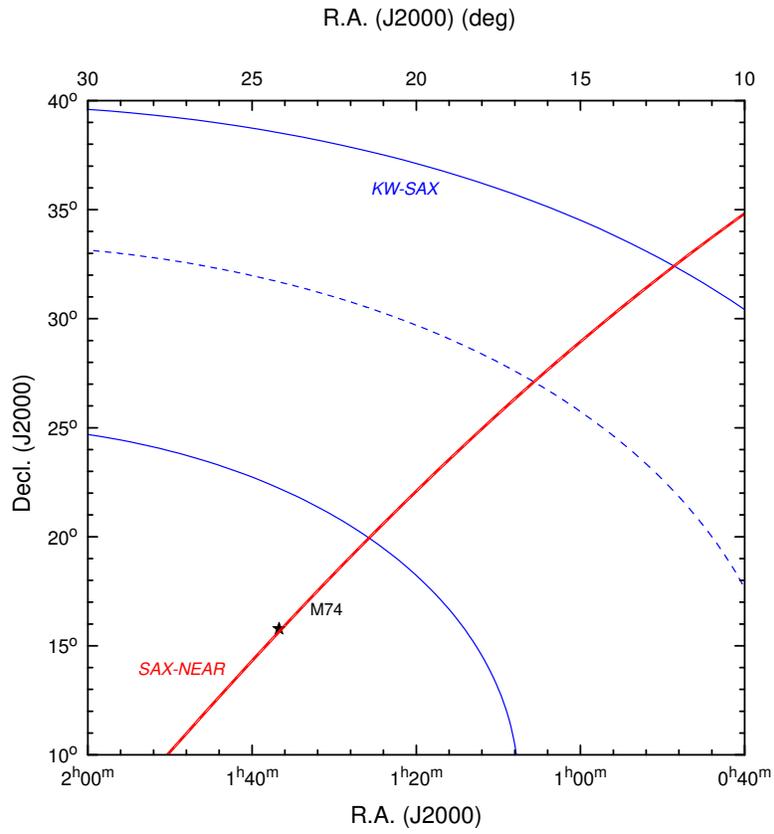}
\caption{IPN localization of GRB 000420 (=GRB20000420\_T42714). The
1.7$\arcmin$ wide \textit{SAX-NEAR} annulus passes through the
nearby M74 galaxy, while the galaxy is well outside the wide
3$\sigma$ \textit{KW-SAX} annulus.}
\label{Fig_GRB000420}
\end{figure}

GRB~000420 (=GRB20000420\_T42271): based on the \textit{KW-NEAR}
annulus it was suggested by \citet{ofek07} that this burst might be
associated with the nearby Sc-type galaxy M74 (NGC 628). The
position of this galaxy lies well outside the wide \textit{KW-SAX}
annulus, thereby excluding it as a possible host for this short GRB
-- see Figure~\ref{Fig_GRB000420}.

GRB~990405 (=GRB19990405\_T30059): initially this event was
classified as a burst from SGR~1900+14 since the narrow
\textit{SAX-Ulysses} annulus (3$\sigma$ half-width of 0$\fdg$035)
passes through the position of this SGR. The derived wide
\textit{KW-SAX} annulus (3$\sigma$ half-width of 6\fdg4) is also
consistent with the SGR position. But this burst is substantially
harder even than two unusually hard bursts from SGR 1900+14: 981022,
991001 \citep{woods99}, making the possible association of this
burst with the SGR doubtful.

\section{SUMMARY AND CONCLUSION}
This paper continues a series of catalogs of gamma-ray burst
localizations obtained by arrival-time analysis, or ``triangulation"
between the spacecraft in the 3rd interplanetary network, as
summarized in Table~\ref{Table_IPNcatalogs}.

We have presented the most comprehensive IPN localization data on
271 Konus-\textit{Wind} short bursts. For 254 bursts IPN error
regions were obtained and for 17 bursts precisely localized by
instruments with imaging capability IPN triangulation annuli were
derived for calibration purposes.

In total we derived 517 triangulation annuli, including 150 annuli
with distant s/c.

It was shown that for many shorts bursts \textit{KW}--near-Earth s/c
(or \textit{INTEGRAL}) triangulation yields a rather narrow annulus
(with half-width sometimes comparable to or even better than the
annuli using distant s/c data), thereby providing small error boxes
with areas of several hundred arcmin$^2$ even for those \textit{KW}
short bursts which were detected by only one distant s/c (and one or
more near-Earth s/c), and providing a long box in cases where the
burst was detected by Konus-\textit{Wind}, \textit{INTEGRAL}
SPI-ACS, and one or more near-Earth s/c.

The localizations can be used for a wide variety of purposes,
including, but not limited to, searches for a) gravitational wave
and neutrino signals from merging compact objects b) very high
energy photons from the burst sources c) giant SGR flares in nearby
galaxies.

\acknowledgements The Konus-\textit{Wind} experiment is supported by
a Russian Space Agency contract and RFBR grants 12-02-00032a and
13-02-12017-ofi-m. K.H. is grateful for IPN support under the
following NASA, JPL, and MIT grants and contracts. JPL 958056 and
1268385 (Ulysses); NNX07AH52G and NNX12AE41G (ADA and ADAP);
NAG5-12614, NNG04GM50G, NNG06GE69G, NNX07AQ22G, NNX08AC90G,
NNX08AX95G and NNX09AR28G (INTEGRAL); NNG05GTF72G, NNG06GI89G,
NNX07AJ65G, NNX08AN23G, NNX09AO97G, NNX10AI23G, and NNX12AD68G
(Swift); NAG5-3500 and NAG5-9503 (NEAR); MIT-SC-R-293291 and
NAG5-11451 (HETE-II); JPL 1282043 (Odyssey); NNX06AI36G, NNX08AB84G,
NNX08AZ85G, NNX09AV61G, NNX10AR12G (Suzaku); NNX09AU03G, NNX10AU34G,
and NNX11AP96G (Fermi); NNX07AR71G (MESSENGER); NAG5-7766,
NAG5-9126, and NAG5-10710 (BeppoSAX).


%

%
\begin{deluxetable}{ccccrcccc}
\tabletypesize{\scriptsize} \tablewidth{0pt}
\rotate
\tablecaption{IPN/Konus short gamma-ray bursts  \label{Table_Basic}}
%
%
\tablehead{
\colhead{Designation}%
& \colhead{Date}%
& \colhead{Konus-\textit{Wind}}%
& \colhead{Type}%
& \colhead{Time delay\tablenotemark{a}}%
& \colhead{Observed by\tablenotemark{b}}%
& \colhead{N$_{tot}$}%
& \colhead{N$_{dist}$}%
& \colhead{Note}\\
\colhead{}%
& \colhead{}%
& \colhead{trigger time (UT)}%
& \colhead{}%
& \colhead{(s)}%
& \colhead{}%
& \colhead{}%
& \colhead{}%
& \colhead{} }
\startdata
GRB19950210\_T08424 & 1995 Feb 10 & 02:20:24.148 & I & -2.602(-0.009,+0.007) & Uly(T),GRO(\#3410) &  3 & 1 & \\
GRB19950211\_T08697 & 1995 Feb 11 & 02:24:57.749 & I &  0.003(-0.009,+0.005) & Uly(T),GRO(\#3412),SRS(T) & 4 & 1 & \\
GRB19950414\_T40882 & 1995 Apr 14 & 11:21:22.798 & I &  0.350(-0.008,+0.006) & Uly(R) & 2 & 1 & \\
GRB19950419\_T08628 & 1995 Apr 19 & 02:23:48.860 & I &  0.418(-0.030,+0.026) & Uly(T) & 2 & 1 & \\
GRB19950523\_T31302 & 1995 May 23 & 08:41:42.284 & I &  0.436(-0.068,+0.040) & Uly(T) & 2 & 1 & \\
\enddata
\tablenotetext{a}{Propagation time delay from \textit{Wind} to the
Earth center and its 3$\sigma$ uncertainty. This delay should be
added to the Konus-\textit{Wind} trigger time to get the
Earth-crossing time.}
\tablenotetext{b}{%
AGI: \textit{Astro-rivelatore Gamma a Immagini LEggero} (AGILE); %
GRO: \textit{Compton Gamma-Ray Observatory (CGRO)}; %
COM: COMPTEL on \textit{CGRO}; %
DMS: \textit{Defense Meteorological Satellite Program}; %
Fer: \textit{Fermi}; %
Hel: Helicon-\textit{Coronas-F}; %
HET: \textit{HETE-2}; %
INT: \textit{International Gamma-Ray Laboratory (INTEGRAL)}; %
KA1: Konus-A on \textit{Cosmos 2326}; %
KA2: Konus-A2 on \textit{Cosmos 2367}; %
KA3: Konus-A3 on \textit{Cosmos 2421}; %
KRF: Konus-RF on \textit{Coronas-Photon}; %
LAT: \textit{Fermi} Large Area Telescope; %
MAXI: \textit{Monitor of All-sky X-ray Image}; %
MES: \textit{Mercury Surface, Space Environment, Geochemistry, and
Ranging} mission (\textit{MESSENGER});
MO: \textit{Mars Odyssey}; %
NEA: \textit{Near Earth Asteroid Rendezvous} mission (\textit{NEAR}); %
RHE: \textit{Ramaty High Energy Solar Spectroscopic Imager (RHESSI)}; %
SAX: \textit{BeppoSAX}; %
SRS: \textit{Stretched Rohini Satellite Series} (SROSS); %
Suz: \textit{Suzaku}; %
Swi: \textit{Swift}; %
TGR: \textit{Transient gamma-ray spectrometer (TGRS)} on \textit{Wind}; %
Uly: \textit{Ulysses}; 
In parentheses the detection mode is given: T -- trigger, R -- rate increase; \#n -- trigger number (when available).}%
\tablenotetext{1}{Imaged by \textit{HETE-2} (WXM and SXC).}
\tablenotetext{2}{Imaged by \textit{Swift}-BAT.}
\tablenotetext{3}{Imaged by \textit{INTEGRAL} IBIS/ISGRI.}
\tablecomments{This table will be published in its entirety in the
electronic edition of the Astrophysical Journal Supplement Series. A
portion is shown here for guidance regarding its form and content.}
\end{deluxetable}

\begin{deluxetable}{cclrrrrr}
\tabletypesize{\scriptsize}
%
%
%
\tablecaption{IPN localization data\label{Table_Annuli}}
%
%
\tablehead{
\colhead{Designation} & N & \colhead{Location \tablenotemark{a}} &
\colhead{R.A.(J2000)} & \colhead{Decl.(J2000)} &
\colhead{$\theta$} & \colhead{$d_{-}(\theta)$} & \colhead{$d_{+}(\theta)$}\\
\colhead{} & \colhead{} & \colhead{source} & \colhead{(deg)} &
\colhead{(deg)} & \colhead{(deg)} & \colhead{(deg)} &
\colhead{(deg)}}
\startdata
GRB19950210\_T08424 & 4 & Uly-GRO & 155.4443 & +25.7475 & 53.6317 & -0.0078 & +0.0078 \\
 & & KW-GRO & 130.3580 & +18.8483 & 52.3229 & -0.1532 & +0.1189 \\
 & & Ecl.Band & 90.000 & -66.561 & 43.3 & -19.5 & +46.0 \\
 & & BATSE & 154.55 & -27.48 & 0 & 0 & 1.15 \\
GRB19950211\_T08697 & 4 & Uly-GRO & 335.8036 & -25.1240 & 85.8298 & -0.0063 & +0.0063 \\
 & & KW-GRO & 311.5868 & -18.2357 & 89.9796 & -0.0678 & +0.1221 \\
 & & Ecl.Band & 270.000 & 66.561 & 65.5 & -39.5 & +22.6 \\
 & & BATSE & 9.51 & 52.65 & 0 & 0 & 1.08\\
\enddata
\tablenotetext{a}{sc$_1$-sc$_2$ -- IPN annulus obtained using sc$_1$
and sc$_2$ data; Ecl.Band -- the ecliptic latitude band of the
burst; sc -- spacecraft/instrument which imaged or autonomously
localized the burst; Occ.sc -- the occulted part of the sky for sc.
The s/c name abbreviations are given in Table~\ref{Table_Basic}. For
autonomous localizations the instrument names are given (i.e.,
BATSE, GRBM, GBM). BATSEbg -- localization from the BATSE
untriggered burst catalog \citep{stern01}. BATSEshift -- BATSE
localization modified as described in Sec.~\ref{Sec_KWBATSEann}.}
\tablecomments{This table will be published in its entirety in the
electronic edition of the Astrophysical Journal Supplement Series. A
portion is shown here for guidance regarding its form and content.}
\end{deluxetable}

\begin{deluxetable}{ccccrrccc}
\tabletypesize{\scriptsize}
\tablewidth{0pt}
\rotate
\tablecaption{IPN error regions\label{Table_Boxes}}
%
%
\tablehead{
\colhead{Designation} & \colhead{$N_{r}$} & \colhead{$N_{c}$} &
\colhead{Type\tablenotemark{a}} &
 \colhead{Max. dim.} & \colhead{Area} & \colhead{Formed
 by\tablenotemark{b}} & \colhead{R.A.(J2000)\tablenotemark{c}} &%
 \colhead{Decl.(J2000)\tablenotemark{c}}\\
\colhead{} & \colhead{} & \colhead{} & \colhead{} & \colhead{(deg)}
& \colhead{(sq. deg)} &\colhead{} & \colhead{(deg)} &
\colhead{(deg)}}
\startdata
GRB19950210\_T08424 & 1 & 4 & B & 5.65E-001 & 8.39E-003 &
Uly-GRO,KW-GRO,Ecl.Band & 154.6820 & -27.8792\\
 & & & & & & & 154.3544 & -27.8662\\
 & & & & & & & 154.9633 & -27.8744\\
 & & & & & & & 154.9330 & -27.8897\\
 & & & & & & & 154.3240 & -27.8812\\
GRB19950211\_T08697 & 1 & 4 & B & 5.13E-001 & 6.09E-003 &
Uly-GRO,KW-GRO,Ecl.Band & 14.7526 & +53.8367\\
 & & & & & & & 14.5265 & +53.9214\\
 & & & & & & & 15.1974 & +53.6449\\
 & & & & & & & 15.1690 & +53.6722\\
 & & & & & & & 14.4975 & +53.9485\\
\enddata
\tablenotetext{a}{Type of the error region: `B' (box), `LB' (long
box: maximum dimension $>10^\circ$),
`S' (annulus segment), and `A' (entire annulus).}%
\tablenotetext{b}{The localization data from Table~\ref{Table_Annuli} used to form the region(s).}%
\tablenotetext{c}{For localizations which are an entire annulus,
these fields are empty (and only one row for the burst is given).
The parameters of the annuli can be found in Table~\ref{Table_Annuli}.}%
%
%
\tablecomments{This table will be published in its entirety in the
electronic edition of the Astrophysical Journal Supplement Series. A
portion is shown here for guidance regarding its form and content.}
\end{deluxetable}

\begin{deluxetable}{ccc}
\tabletypesize{\scriptsize}
\tablecaption{Recent IPN catalogs of gamma-ray
bursts\label{Table_IPNcatalogs}}
\tablewidth{0pt} \tablehead{ \colhead{Years covered} &
\colhead{Number of GRBs} & \colhead{Description} } \startdata

1990--1992 & 16            & \textit{Ulysses, Pioneer Venus Orbiter,} WATCH, SIGMA, PHEBUS GRBs\tablenotemark{a} \\
1990--1994 & 56            & \textit{Granat-}WATCH supplement\tablenotemark{b} \\
1991--1992 & 37            & \textit{Pioneer Venus Orbiter, Compton Gamma-Ray Observatory, Ulysses} GRBs\tablenotemark{c} \\
1991--1994 & 218           & BATSE 3B supplement\tablenotemark{d} \\
1991--2000 & 211           & BATSE untriggered burst supplement\tablenotemark{e} \\
1992--1993 & 9             & \textit{Mars Observer} GRBs\tablenotemark{f} \\
1994--1996 & 147           & BATSE 4Br supplement\tablenotemark{g} \\
1994--2010 & 279           & Konus short bursts\tablenotemark{h} \\
1996--2000 & 343           & BATSE 5B supplement\tablenotemark{i} \\
1996--2002 & 475           & \textit{BeppoSAX} supplement\tablenotemark{j} \\
2000--2006 & 226           & HETE-2 supplement\tablenotemark{k} \\
2008--2010 & 146           & GBM supplement\tablenotemark{l}
\enddata
\tablenotetext{a}{\citet{hurley00a}}
\tablenotetext{b}{\citet{hurley00c}}
\tablenotetext{c}{\citet{laros98}}
\tablenotetext{d}{\citet{hurley99a}}
\tablenotetext{e}{\citet{hurley05}}
\tablenotetext{f}{\citet{laros97}}
\tablenotetext{g}{\citet{hurley99b}}
\tablenotetext{h}{present catalog}
\tablenotetext{i}{\citet{hurley11a}}
\tablenotetext{j}{\citet{hurley10a}}
\tablenotetext{k}{\citet{hurley11b}}
\tablenotetext{l}{\citet{hurley13}}
\end{deluxetable}



\begin{thebibliography}{}
%
\bibitem[Abadie et al. (2012)]{abadie12} Abadie, J., Abbott, B. P., Abbott, T. D., et al. 2012, \apj \, 755, 2
%
\bibitem[Abbott et al. (2008)]{abbott08} Abbott, B., Abbott, R., Adhikari, R., et al., 2008, \apj \, 681, 1419
%
\bibitem[Arimoto et al.(2006)]{arimoto06}Arimoto, M., Ricker, G., Atteia, J-L., et al. 2006, GCN Circ. 4550
%
\bibitem[Aptekar et al.(1995)]{aptekar95} Aptekar, R., Frederiks, D., Golenetskii, S., et al. 1995, \ssr \, 71, 265
%
\bibitem[Aptekar et al.(1998)]{aptekar98} Aptekar, R. L., Butterworth, P. S., Cline, T. L., et al. 1998, \apj \, 493, 404
%
\bibitem[Atteia et al.(2003)]{atteia03} Atteia, J.-L., Boer, M., Cotin, F., et al., 2003, in Gamma-Ray Burst and Afterglow Astronomy 2001, A Workshop Celebrating the First Year of the HETE Mission, Eds. G. Ricker and R. Vanderspek, AIP Conf. Proc. 662 (AIP: New York), 17
%
\bibitem[Barthelmy et al.(2005)]{barthelmy05} Barthelmy, S. D., Barbier, L. M., Cummings, J. R., et al. 2005, \ssr \, 120, 143
%
\bibitem[Boynton et al.(2004)]{boynton04} Boynton, W. V., Feldman, W. C., Mitrofanov, I. G., et al. 2004, \ssr \, 110, 37
%
\bibitem[Briggs et al. (1999)]{briggs99} Briggs, M. S., Pendleton, G. N., Kippen, R. M., et al. 1999, \apjs \, 122, 503
%
\bibitem[Feroci et al.(1997)]{feroci97} Feroci, M., Frontera, F., Costa, E., et al., 1997, in EUV, X-Ray, and Gamma-Ray Instrumentation for Astronomy VIII, Eds. O. Siegmund and M. Gummin, SPIE 3114, 186
%
\bibitem[Frederiks et al.(2007)]{frederiks07} Frederiks, D. D., Palshin, V. D., Aptekar, R. L., et al. 2007, Astronomy Lettrs, 33, 19
%
\bibitem[Fishman et al.(1992)]{fishman92} Fishman, G., Meegan, C., Wilson, R., Paciesas, W., and Pendleton, G., 1992, in Proc. Compton Observatory Science Workshop, ed. C. Shrader, N. Gehrels, and B. Dennis (NASA CP 3137; Greenbelt, MD: GSFS), 26
%
\bibitem[Frontera et al.(1997)]{frontera97} Frontera, F., Costa, E., dal Fiume, D., et al. 1997, Astron. Astrophys. Suppl. Ser. 122, 357
%
\bibitem[Frontera et al.(2009)]{frontera09} Frontera, F., Guidorzi, C., Montanari, E., et al. 2009, \apjs \, 180, 192
%
\bibitem[Gehrels et al. (2004)]{gehrels04} Gehrels, N., Chincarini, G., Giommi, P., et al. 2004, \apj \, 611, 1005
%
\bibitem[Gold et al.(2001)]{gold01} Gold, R., Solomon, S., McNutt, R., et al. 2001, \planss, 49, 1467
%
\bibitem[Golenetskii et al.(1974)]{golenetskii74} Golenetskii, S. V., Il'Inskii, V. N., \& Mazets, E. P.
1974, Cosmic Research, 12, 706
%
\bibitem[Goldsten et al.(2007)]{goldsten07} Goldsten, J. O., Rhodes, E. A., Boynton, W. V., et al. 2007, \ssr \, 131, 339
%
\bibitem[G\"{o}tz et al.(2007)]{gotz07} G\"{o}tz, D., Beckmann, V., Mereghetti, S., \&
Paizis, A. 2007, GCN Circ. 6608
%
\bibitem[Hurley et al.(1992)]{hurley92} Hurley, K., Sommer, M., Atteia, J.-L., et al. 1992, Astron. Astrophys. Suppl. Ser. 92(2), 401
%
\bibitem[Hurley et al.(1999a)]{hurley99a} Hurley, K., Briggs, M., Kippen, R. M., et al. 1999a, \apjs \, 120, 399
%
\bibitem[Hurley et al.(1999b)]{hurley99b} Hurley, K., Briggs, M., Kippen, R. M., et al. 1999b, \apjs \, 122, 497
%
\bibitem[Hurley et al.(2000a)]{hurley00a} Hurley, K., Laros, J., Brandt, S., et al. 2000a, \apj \, 533, 884
%
%
\bibitem[Hurley et al.(2000c)]{hurley00c} Hurley, K., Lund, N., Brandt, S., et al. 2000c, \apjs \, 128, 549
%
\bibitem[Hurley et al.(2005)]{hurley05} Hurley, K., Stern, B., Kommers, J., et al. 2005, \apjs \, 156, 217
%
\bibitem[Hurley et al.(2006)]{hurley06} Hurley, K., Mitrofanov, I., Kozyrev, A., et al. 2006, \apjs \, 164, 124
%
\bibitem[Hurley et al.(2010a)]{hurley10a} Hurley, K., Guidorzi, C., Frontera, F., et al. 2010, \apjs \, 191, 179
%
\bibitem[Hurley et al.(2010b)]{hurley10b} Hurley, K., Rowlinson, A., Bellm, E., et al. 2010, \mnras \, 403, 342
%
\bibitem[Hurley et al.(2011a)]{hurley11a} Hurley, K., Briggs, M., Kippen, R. M., et al. 2011, \apjs \, 196, 1
%
\bibitem[Hurley et al.(2011b)]{hurley11b} Hurley, K., Atteia, J.-L., Barraud, C., et al. 2011, \apjs \, 197, 34
%
\bibitem[Hurley et al.(2013)]{hurley13} Hurley, K., Pal'shin, V., Aptekar, R., et al. 2013, \apjs, 207, 39
%
\bibitem[von Kienlin et al. (2003)]{kienlin03} von Kienlin, A., Beckmann, V., Rau, A., et al. 2003, \aap, 411, L299
%
\bibitem[Kommers et al.(2000)]{kommers00} Kommers, J., Lewin, W., Kouveliotou, C., van Paradijs, J., Pendleton, G., Meegan, C., and Fishman, G., 2000, \apj \, 533, 696
%
%
\bibitem[Laros et al.(1997)]{laros97} Laros, J., Boynton, W., Hurley, K., et al. 1997, \apjs \, 110, 157
%
%
\bibitem[Laros et al.(1998)]{laros98} Laros, J., Hurley, K., Fenimore, E., et al. 1998, \apjs \, 118, 391
%
\bibitem[Lichti et al.(2000)]{lichti00} Lichti, G. G., Georgii, R., von Kienlin, A., et al. 2000, in The Fifth Compton Symposium, Eds. Mark L. McConnell \& James M. Ryan, AIP Conf. Proc. 510  (AIP: New York), p. 722
%
\bibitem[Lin et al.(2002)]{lin02} Lin, R. P., Dennis, B. R., Hurford, G. J., et al. 2002, \solphys, 210, 3
%
\bibitem[Marar et al.(1994)]{marar94} Marar, T. M. K., Sharma, M. R., Seetha, S., et al., 1994, \aap \, 283, 698
%
\bibitem[Mazets \& Golenetskii (1981)]{mazets81} Mazets, E. P., \& Golenetskii, S. V., 1981, Astrophys. Space Sci. 75, 47
%
\bibitem[Mazets et al.(2008)]{mazets08} Mazets, E. P., Aptekar, R. L., Cline, T. L., et al., 2008, \apj \, 680, 545
%
\bibitem[Meegan et al.(2009)]{meegan09} Meegan, C., Lichti, G., Bhat, P. N., et al., 2009, \apj \, 702, 791
%
\bibitem[Norris, Scargle, \& Bonnell (2001)]{norris01}Norris, J. P.,  Scargle, J. D., \& Bonnell, J. T. 2001, in Gamma-Ray Bursts in the Afterglow Era, ed. E. Costa,
F. Frontera, \& J. Hjorth (Berlin: Springer), 40
%
%
\bibitem[Ofek et al. (2006)]{ofek06} Ofek, E. O., Kulkarni, S. R., Nakar, E., et al.  2006, \apj \, 652, 507
%
\bibitem[Ofek (2007)]{ofek07} Ofek, E. O. 2007, \apj \, 659, 339
%
\bibitem[Ofek et al. (2008)]{ofek08} Ofek, E. O., Muno, M., Quimby, R., et al., 2008, \apj \, 681, 1464
%
\bibitem[Oraevskii et al. (2002)]{oraevskii02} Oraevskii, V. N.,  Sobelman, I. I., Zhitnik, I. A., \& Kuznetsov, V. D.,
2002, Phys. Usp. 45, 886
%
\bibitem[Paciesas et al. (2012)]{paciesas12} Paciesas, W. S., Meegan, C., von Kienlin, A., et al. 2012, \apjs, 199, 18
%
\bibitem[Rau et al. (2004)]{rau04} Rau, A., von Kienlin, A., Lichti, G., Hurley, K., \& Beck, M. 2004, GCN Circ. 2568
%
\bibitem[Rau et al. (2005)]{rau05} Rau, A., von Kienlin, A., Hurley, K., and Lichti, G. 2005, \aap \, 438, 1175
%
\bibitem[Ricker et al.(2003)]{ricker03} Ricker, G. R., Atteia, J.-L., Crew, G. B., et al., 2003, in Gamma-Ray Burst and Afterglow Astronomy 2001, A Workshop Celebrating the First Year of the HETE Mission, Eds. G. Ricker and R. Vanderspek, AIP Conf. Proc. 662 (AIP: New York), 3, 2003
%
\bibitem[Sakamoto et al.(2011)]{sakamoto11} Sakamoto, T., Barthelmy, S. D., Baumgartner, W. H., et al., 2011, \apjs \, 195, 2
%
\bibitem[Saunders et al.(2004)]{saunders04} Saunders, R. S., Arvidson, R. E., Badhwar, G. D, et al., 2004, \ssr \, 110, 1
%
%
\bibitem[Smith et al.(2002)]{smith02} Smith, D. M., Lin, R., Turin, P., et al. 2002, \solphys \, 210, 33
%
\bibitem[Solomon et al.(2007)]{solomon07} Solomon, S. C., McNutt, R. L., Gold, R. E., \& Domingue, D. L. 2007, \ssr \, 131, 3
%
\bibitem[Stern et al.(2001)]{stern01} Stern, B., Tikhomirova, Y., Kompaneets, D., Svensson, R., and Poutanen, J., 2001, \apj \, 563, 80
%
\bibitem[Tavani et al.(2009)]{tavani09} Tavani, M., Barbiellini, G., Argan, A., et al., 2009, \aap \, 502, 995
%
\bibitem[Takahashi et al.(2007)]{takahashi07} Takahashi, T., Abe, K., Endo, M., et al., 2007, \pasj \, 59, S35
%
\bibitem[Terrell et al.(1996)]{terrell96} Terrell, J., Lee, P., Klebesadel, R., \& Griffee, 1996, in 3rd Huntsville Symposium, AIP Conf. Proc. 384 (AIP: New York), Eds. C. Kouveliotou, M. Briggs, and G. Fishman, 545
%
\bibitem[Terrell et al.(1998)]{terrell98} Terrell, J., Lee, P., Klebesadel, R., \& Griffee, J., 1998, in Gamma-Ray Bursts, 4th Huntsville Symposium, Eds. C. Meegan, R. Preece, and T. Koshut, AIP Conf. Proc. 428, AIP Press (New York), p. 54
%
\bibitem[Terrell et al.(2004)]{terrell04} Terrell, J., \& Klebesadel, R., 2004 , in Gamma-Ray Bursts: 30 Years of Discovery, Eds. E. Fenimore and M. Galassi, AIP Conf. Proc. 727 (AIP: New York), p. 541
%
\bibitem[Trombka et al.(1999)]{trombka99} Trombka, J. I., Boynton, W. V., Br\"{u}ckner, J., et al., 1999, Nucl. Inst. And Methods in Physics Research A 422, 572
%
\bibitem[Yamaoka et al.(2009)]{yamaoka09} Yamaoka, K., Endo, A., Enoto, T., et al., 2009, \pasj \, 61, S35
%
\bibitem[Woods et al.(1999)]{woods99} Woods, P. M., Kouveliotou, C., van Paradijs, J., et al., 1999, \apj, 527, L47
%
%
\bibitem[Zhang et al.(2009)]{zhang09} Zhang, B., Zhang, B.-B., Virgili, F., et
al., 2009, \apj \, 703, 1696
%
\bibitem[Zhang et al.(2010)]{zhang10} Zhang, X.-L., Rau, A., \& von Kienlin, A. 2010, in proceeding of the 8th INTEGRAL Workshop The Restless Gamma-ray Universe, 2010 September 27–30, Dublin, Ireland, PoS (INTEGRAL 2010)161
%
\end{thebibliography}
\end{document}